\documentclass[rmp,twocolumn,superscriptaddress,showpacs,floatfix]{revtex4}
\usepackage{amsmath}
\usepackage{graphicx}

\begin{document}

\title{{\it Colloquium:} Phononics: Manipulating heat flow with electronic analogs and beyond}

\author{Nianbei Li}
\affiliation{Center for Phononics and Thermal Energy Science, Department of Physics, Tongji University, 200092 Shanghai, P.R. China}
\affiliation{Department of Physics and Centre for Computational
Science and Engineering,
 National University of Singapore, Singapore 117546}
\affiliation{Max Planck Institute for the
Physics of Complex Systems, N\"othnitzer Strasse 38, D-01187
Dresden, Germany}
\author{Jie Ren}
\affiliation{Department of Physics and Centre for Computational
Science and Engineering,
 National University of Singapore, Singapore 117546}
\affiliation{NUS Graduate School for Integrative Sciences and
Engineering, National University of Singapore, 117456 Singapore}
\author{Lei Wang}
\affiliation{Department of Physics, Renmin University of China,
Beijing 100872, P. R. China}
\affiliation{Department of Physics and
Centre for Computational Science and Engineering,
 National University of Singapore, Singapore 117546}
\author{Gang Zhang}
\affiliation{Department of Electronics, Peking University, Beijing
100872, P. R. China} \affiliation{Department of Physics and Centre for
Computational Science and Engineering,
 National University of Singapore, Singapore 117546}
\author{Peter H\"anggi}
\affiliation{Department of Physics and Centre for Computational
Science and Engineering,
National University of Singapore, Singapore 117546}
\affiliation{Max Planck Institute for the
Physics of Complex Systems, N\"othnitzer Strasse 38, D-01187
Dresden, Germany}
\affiliation{Institut f\"ur Physik, Universit\"at Augsburg,
Universit\"atsstrasse 1, D-86135 Augsburg, Germany}
\author{Baowen Li}
\email{phylibw@nus.edu.sg}
\affiliation{Department of Physics and Centre for Computational
Science and Engineering,
National University of Singapore, Singapore 117546}
\affiliation{NUS Graduate School for Integrative Sciences and
Engineering, National University of Singapore, 117456 Singapore}
\affiliation{Center for Phononics and Thermal Energy Science,
Department of Physics, Tongji University, 200092 Shanghai, P.R. China}

\date{\today}

\begin{abstract}
The form of energy termed heat that typically derives from  lattice
vibrations, i.e.  the phonons, is usually considered as  waste
energy and, moreover, deleterious to information processing.
However, with this colloquium, we attempt to rebut this common view:
By use of tailored models we demonstrate that phonons can be
manipulated like electrons and photons can, thus enabling controlled
heat transport. Moreover, we  explain that phonons can be put to
beneficial use  to carry and process information. In a first part we
present ways to control heat transport and how to process
information  for physical systems which are driven by a temperature bias.
Particularly, we put forward the toolkit of familiar electronic analogs for exercising
phononics; i.e. phononic devices which act as thermal
diodes, thermal transistors, thermal logic gates and thermal
memories, etc.. These concepts are then put to work to
transport, control and rectify heat in physical realistic nanosystems by
devising practical designs of hybrid nanostructures that permit
the operation of functional
phononic devices and, as well, report  first experimental realizations.
Next, we discuss yet richer possibilities to manipulate  heat flow by use of time varying thermal
bath temperatures or various other external fields. These give rise to a plenty of intriguing
phononic nonequilibrium phenomena as for example the directed shuttling of heat,
a  geometrical phase induced heat pumping, or the phonon Hall effect,
that all may find its way into operation with electronic analogs.  \\
\end{abstract}


\maketitle
\tableofcontents



\section{Introduction} \label{Sec I}
When it comes to the task of transferring energy, nature has at its disposal  tools such as electromagnetic radiation,  conduction by electricity and also heat. The latter two tools play a dominant role  from a technological viewpoint. The  conduction of heat and electric conduction are, however, two fundamental energy transport mechanisms of comparable importance,  although never been treated equally in science. Modern information processing rests on  micro electronics, which after the  invention of the electronic solid-state transistor \cite{Bardeen1948PR74}, and other related devices, sparked off an unparalleled technological development, the state of the art being electric integrated circuitry.  Without any doubt,  this technology markedly changed many aspects of our daily life. Unfortunately, a similar technology which builds on electronic analogs via the constructive use of heat flow has not yet been realized by mankind, although several attempts have been repeatedly undertaken. In everyday life, however,  signals encoded by heat prevail over those by electricity. Therefore, the potential of using heat control may result in an even more abundant and unforseen wealth of applications. A legitimate question then is: Does phononics, i.e., the counterpart technology of  electronics present only a dream?

Admittedly, it indeed is substantially more difficult to control {\it a priori} the flow of heat in a solid than it is to control the flow of electrons. The source of this imbalance is that,  unlike electrons, the carriers of heat --  the phonons -- are quasi-particles in the form of  energy bundles that possess neither a bare mass nor a bare charge.  Although isolated phonons by itself do not influence each other, interactions involving phonons become of importance in the presence  of condensed phases.  Some examples that come to mind are phonon polaritons, i.e. the interaction of optical phonons with infrared photons, the  generic  phonon-electron interactions occurring in metals and semi-metal structures, phonon-spin interactions, or the phonon-phonon interaction in presence of nonlinearity. Therefore, heat flow features aspects which in many  ways are distinct from charge and matter  flow. Nonetheless, there occur in condensed phases many interesting cross-interplays as  it is encoded with the reciprocal relations of the Onsager form for mass, heat and charge flow, of which thermoelectricity \cite{Callen1960book,Dubi2011RMP83} or thermophoresis, i.e. the Soret {\it vs.} its reciprocal Dufour effect \cite{Callen1960book},  are typical exemplars. Therefore, capitalizing on the rich physical diversities involving  phonon transport as obtained with recent successes in  nano-technology may open the door to turn  phononics from a dream into a reality.

With this Colloquium we shall focus on  two fundamental issues of phononics; - i.e. the manipulation of heat energy flow on the  nanoscale and the objective of processing information by utilizing  phonons. More precisely, we shall investigate the possibilities to devise elementary building blocks for doing phononics; namely we study the conceptual realization and its possible operation  of a  thermal diode which rectifies heat current, a thermal transistor that is capable to switch and amplify heat flow and last but not least a thermal memory device.

The objective of controlling heat flow on the nanoscale necessarily rests on the microscopic laws that govern heat conduction, stability aspects or thermometric issues. For these latter themes the literature provides the readers with several comprehensive reviews and features already. For heat flow and/or related thermoelectric phenomena on the micro-/nano- and molecular scales  we refer the readers to the excellent treatises by \textcite{Lepri2003PR377}, \textcite{Casati2005Chaos15}, \textcite{Galperin2007JPCM19}, \textcite{Dhar2008AdvPhys57}, \textcite{li2005CHAOS15},
\textcite{zhang2010NS2}, \textcite{Pop2010NR3}, and recently  also by \textcite{Dubi2011RMP83}. We further demarcate our presentation from the subjects  of refrigeration on mesoscopic scales and thermometry \cite{Giazotto2006RMP78} and, as well,  as our title suggests, also do not address  {\it per se} in greater detail the issue of conventional way of manipulating heat flow upon changing thermal conductivity by means of various phonon scattering mechanisms in nano- and heterostructures \cite{Ziman1960book, Chen2005book}. For compelling recent developments and advances in this latter area we refer the interested readers to the recent surveys by \textcite{Balandin2005JNN5} and \textcite{Balandin2007JNO2}, together with the original literature cited therein.

\begin{figure}[htbp]
\centerline{\includegraphics[width=7cm]{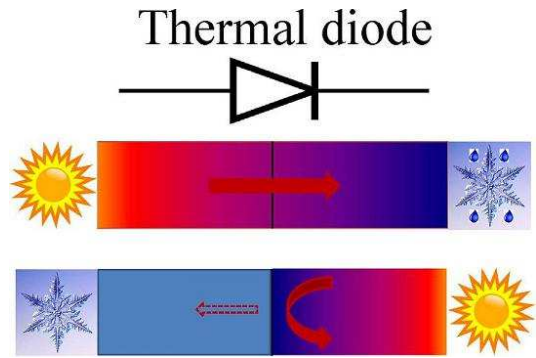}}
\caption{\label{fig:intro_diode}(color).
Sketch of the {\it modus operandi} of a thermal diode. When the left end of the diode is at a higher temperature as compared to its right counterpart  heat is allowed  to flow almost freely. In contrast, when the right end is made hotter in reference to the left end the transduction of heat becomes strongly diminished.}
\end{figure}

In this spirit, a primary building block for doing phononics is a setup that rectifies heat flow; i.e. a thermal rectifier/diode. Such a device acts as a thermal conductor if a positive thermal bias is applied while in the opposite case of a negative thermal bias it undergoes poor thermal conduction, thus effectively acting as a thermal insulator, or possibly also vice versa. The concept of such a thermal diode is sketched in Fig.~ \ref{fig:intro_diode}.

The concept of a thermal diode involves, just as in its electronic counterpart, the presence of a symmetry breaking mechanism.
This symmetry breaking is most conveniently realized by merging two materials exhibiting different heat transport characteristics. Historically, \textcite{Starr1936JAP7} working at Rensselaer Polytechnic Institute in New York built a junction composed of a metallic copper part  which he joined with its cuprous oxide phase; thus proving the working principle of rectifying heat in such a structure. Starr's thermal rectifier is physically based on an asymmetric electron-phonon interaction occurring in the  interface of the two dissimilar materials.  There exist a plenty of such macroscopic rectifiers which function via the  difference of the material response due to  temperature bias and/or other externally applied control  fields such as strain, etc., \cite{Roberts2011IJTS50}.

\begin{figure}[htbp]
\centerline{\includegraphics[width=7cm]{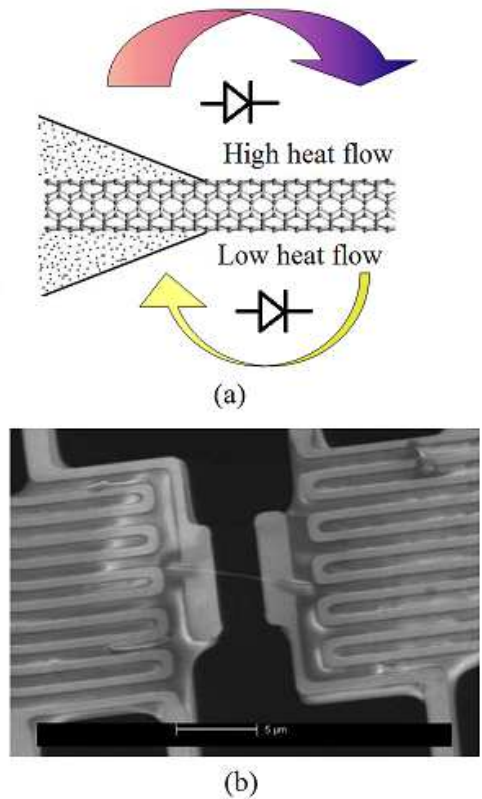}}
\caption{(color online).
Concept and experimental setup of a nanoscale thermal diode. An asymmetric nanotube is placed over two electrodes with one serving as a heater. For further details on the experimental procedures and the experimental findings we refer the readers to the  experiment by \textcite{Chang2006Sci314} and the detailed discussion in Sect. III below. Adapted from \textcite{Chang2006Sci314}.}
\label{fig:intro_exp}
\end{figure}

The focus in this colloquium will be  on a thermal rectification scenario that is induced by phonon transport occurring on the nanoscale.  The concept of  such  a thermal rectifier for heat was put forward by \textcite{terrano2002PRL88}. The authors therein proposed to use a three-segment structure composed of different nonlinear lattice segments.  The underlying physical mechanism relies on the resonance phenomenon for the temperature dependent power spectrum {\it vs.} frequency as a result of  the nonlinear lattice dynamics. Subsequently, it has been shown that a modified two-segment setup \cite{li2004PRL93, li2005PRL95} yields considerably improved rectification characteristics as compared to the original three-segment setup \cite{terrano2002PRL88}. These pioneering works in turn ignited a flurry of activities, manifesting different advantageous features and characteristics.  The theoretical and numerical efforts culminated in a first experimental validation of such a thermal rectifier in 2006: The device itself is based on an asymmetric nanotube structure \cite{Chang2006Sci314}.  The concept of this latter thermal diode together with its explicit experimental setup is depicted with Fig. \ref{fig:intro_exp}.

The thermal diode presents an important first step towards phononics. For performing logic operations and useful circuitry, however, additional control mechanisms for heat are required. This comprises the task of devising (i) the thermal analog of  an electronic transistor, (ii) thermal logic gates  and, as well, (iii) a thermal memory. The physical concept of these salient  phononic building blocks will be elucidated in Sect. II. The main physical feature that enters the function of such phonon devices is the occurrence of {\it negative differential thermal resistance} (NDTR); the latter being a direct consequence of the  inherently acting  nonlinear dynamics with its intriguing nonlinear response to an externally  applied thermal bias.

With Sect. III we investigate how to put these thermal phononic concepts to ``action'' by using realistic  nanoscale  structures. The actual operations of such devices rest on extended molecular dynamics simulations which serve as a guide for implementing its experimental realization. The control of heat flow in the above mentioned phononic building blocks is managed mainly by applying a {\it static} thermal bias. More intriguing control of transport emerges when the manipulations are made explicitly time-dependent or by use of different external forces such as the application of  magnetic fields. As detailed with Sec. IV such  manipulation scenarios then generate new roadways towards fine-tuned control and counterintuitive response behaviors. Originally, such dynamic control has been implemented for anomalous particle transport  by taking the system dynamics out-of-equilibrium: Doing so results in intriguing phenomena such as  Brownian motor (ratchet-like) transport, absolute negative mobility and alike \cite{astumian2002PhysicsToday55, hanggi2009RMP81, hanggi2005AnnalsPhysics14}.  A similar reasoning can be put to work for shuttling {\it heat} in appropriately designed lattice structures, as detailed in  Sect.  IV. In Sect. V, we summarize our main findings, discuss yet some additional elements for phononic concepts and  reflect on future potential  and  visions to advance the field of phononics from its present infancy towards a  mature level.



\section{Phononics Devices: Theoretical Concepts} \label{Sec II}

\subsection{Thermal diode: Rectification of heat flow}
The task of directing heat for information processing  as in electronics requires
a toolkit with suitable building blocks, namely those nonlinear components that mimic the
role of diodes, transistors, and alike, known from electronic circuitry.
A first challenge then is to design the blueprints
for such components that function for heat control analogous to the
building blocks for electronics. This objective
is best approached  by making use of the nonlinear
dynamics present in anharmonic lattice structures in
combination with the implementation of a system inherent symmetry breaking.  We
start with the discussion of the theoretical design for
thermal diodes that rectify heat flow.

\subsubsection{Two-segment thermal diode}\label{II-A-1}

In order to achieve  thermal rectification, we exploit the nonlinear response mechanism as it derives from
inherent temperature-dependent power spectra.
An everyday analog of such a nonlinear frequency response is a playground swing when
driven into its large amplitude regime via parametric resonance. The response is optimized whenever the
natural frequencies match those of the perturbations. Likewise, energy can be transported across
two different segments when the corresponding vibrational frequency response characteristics overlap.

More precisely, whenever the
power spectrum in one part of the device matches with its neighboring part,
we find that heat energy is exchanged efficiently. In the absence of such overlapping spectral properties, the exchange of energy becomes strongly diminished.
Particularly, the response behavior of realistic materials is typically anharmonic by nature.
As a consequence, the corresponding power spectra become strongly dependent on temperature, see  in  Appendix A, Subsect. $3$.
If an asymmetric system is composed of
different parts with differing physical parameters the resulting temperature-dependence of the power
spectra will differ likewise.

\begin{figure}[!htbp]
\includegraphics[width=\columnwidth]{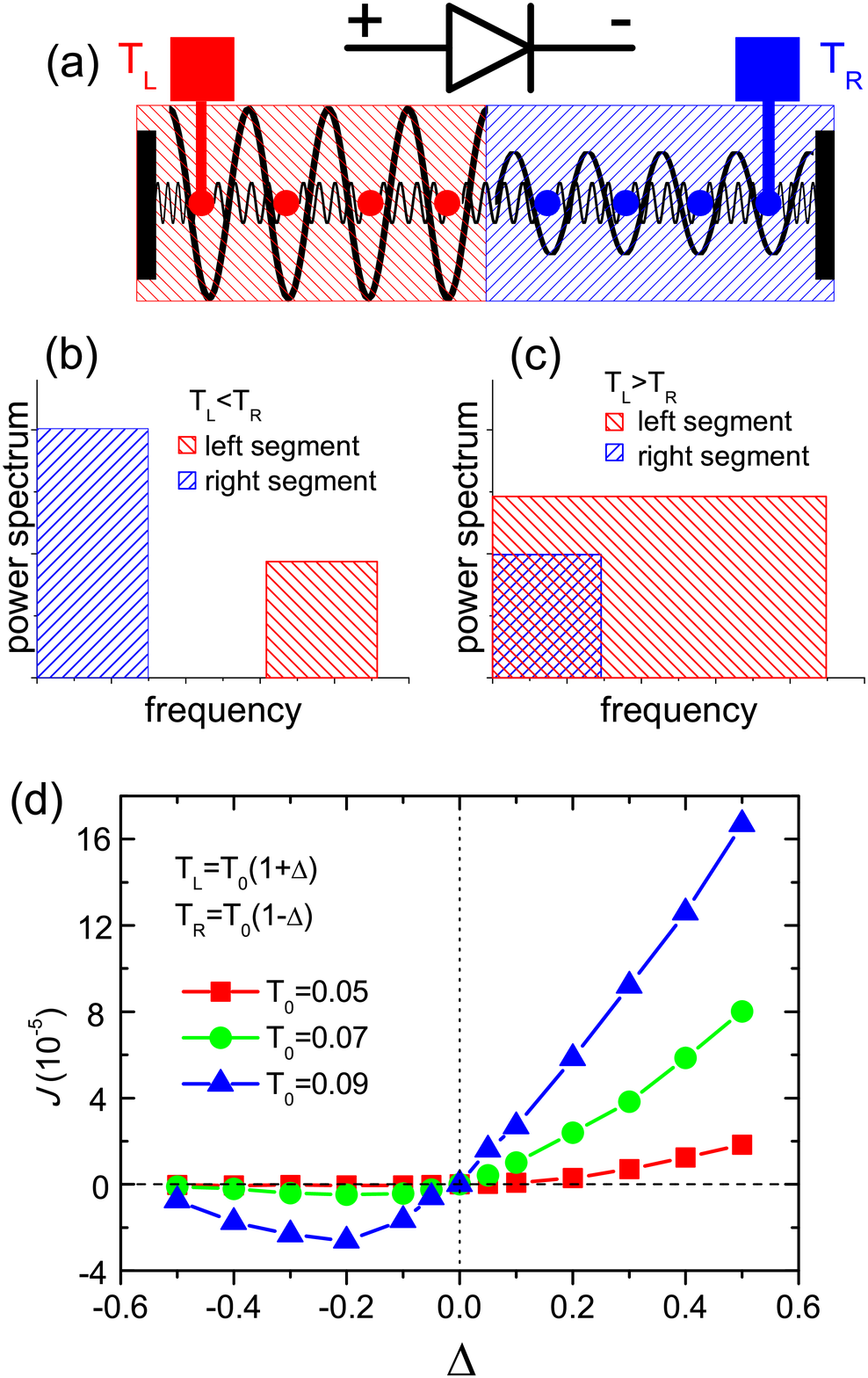}
\vspace{-0.2cm}
\caption{\label{fig:jdelta}(color online). Concept of a thermal diode. Panel (a). Blueprint of an efficient two-segment thermal diode composed of two different Frenkel-Kontorova  chains.  The left (red) segment consisting of a chain of particles subjected to a strong cosinusoidal varying on-site potential (illustrated by the large wavy curve) is connected to the right (blue) segment possessing a relatively weak on-site potential.
(b) In the case that the temperature $T_{\text L}$ in
the left segment is colder than the corresponding right temperature $T_{\text R}$, i.e. $T_{\text L} < T_{\text R}$,
the power spectrum of the particle motions of the left segment
is weighted at high frequencies. This is so because of the difficulty experienced by the dwelling particles to overcome the large barriers of the on-site potential. In contrast, the  power spectrum of the right segment is weighted  at low frequencies.
As a result, the overlap of the spectra is weak, implying that the heat current $J$ becomes strongly diminished.
(c) Here the situation is opposite to panel (b). With $T_{\text L} > T_{\text R}$  the particles can now move almost freely between neighboring
barriers. Consequently the power spectrum  extends to much lower frequencies, yielding an appreciable overlap with the right placed segment. This in turn causes a sizable heat current.
Panel (d). Heat current $J$ {\it vs.} the relative temperature bias  $\Delta$,  as defined in the inset,
for three different values of the reference temperature $T_{\text 0}$.
Adapted from \textcite{wang2008PhysicsWorld21} and \textcite{li2004PRL93}.
}
\end{figure}

Based on these insights, a possible working principle of a thermal diode goes as follows:
If a temperature bias makes the spectral features of different parts overlap with each other, we obtain a favorable energy exchange.
In contrast, if for the opposite temperature bias the spectral
properties of the different parts fail to overlap appreciably  a strong suppression of heat transfer occurs.
In summary, this match/mismatch of spectral properties  provides the salient mechanism  for thermal rectification,
see in Fig.~\ref{fig:jdelta}(b) and (c).

Because the power spectra of an arbitrary nonlinear material typically become temperature-dependent,
the use of any asymmetric nonlinear system is expected to display an inequivalent heat transport upon reversal of the temperature bias.
It is, however, not a simple task to design a device that results in physical designated and  technologically feasible thermal diode properties.
After having investigated a series of possible setups we designed a thermal diode model that
performs efficiently over a  wide range of system parameters \cite{li2004PRL93}.
The blueprint of this device consists of two nonlinear segments which are weakly coupled by a linear spring with strength $k_{\text{int}}$.
Each segment is composed of a chain of particles in which each individual particle is coupled with its nearest neighbors by linear springs. This whole nonlinear two-segment chain is each subjected to a cosinusoidal varying on-site potential; the latter is provided by the coupling to a substrate. These individual chains are therefore described by a Frenkel-Kontorova (FK) lattice dynamics, cf. Appendix. The scheme of this thermal diode is depicted in Fig.~\ref{fig:jdelta}(a).

The key feature of this FK-diode setup is the chosen difference in the strength of the corresponding on-site potential.
At low temperature, the particles are
confined in the valleys of the on-site potential. Thus the power spectrum is  weighted in the high frequency regime.
At high temperatures  the particles assume sufficiently large
kinetic energies so that thermal activation \cite{Hanggi1990RMP62} across the inhibiting barriers becomes feasible. The corresponding
power spectrum is then moved towards lower frequencies. By setting the strength of the on-site potential in the two segments at distinct different levels, see Fig.~\ref{fig:jdelta}(a), we achieve the desired strong thermal rectification. Note that the barrier height of the on-site potential for the right segment is chosen sufficiently small so that the corresponding particles are allowed to  move almost freely, both in the low and in the high temperature regime. In the case that the left end is set at the low temperature, its power spectrum is weighted within the high frequency regime. This in turn causes an appreciable mismatch with  the right segment, see Fig.~\ref{fig:jdelta}(b). In the opposite case, when the left end is set at the high temperature,
its weighted power spectrum  moves towards lower frequencies, thus matching considerably with  the right segment, see Fig.~\ref{fig:jdelta}(c).

In Fig.~\ref{fig:jdelta}(d) the resulting stationary heat current $J$  (see in Appendix A) {\it vs.} the relative temperature bias $\Delta$ is depicted for three values of the reference temperatures $T_0$  \cite{li2004PRL93}. The relation between the dimensionless temperature and the actual physical temperature can be found in the Appendix as well. It is shown that when $\Delta>0$ (i.e. $T_L>T_R$), the heat current gradually increases with increasing $\Delta$, i.e. the setup behaves as a `good' thermal conductor; in contrast,
when $\Delta<0$ ($T_L<T_R$), the heat current remains small. The two-segment structure
thus behaves  as a `poor' thermal conductor, i.e. it mimics a thermal insulator.

\begin{figure}[htbp]
\includegraphics[width=\columnwidth]{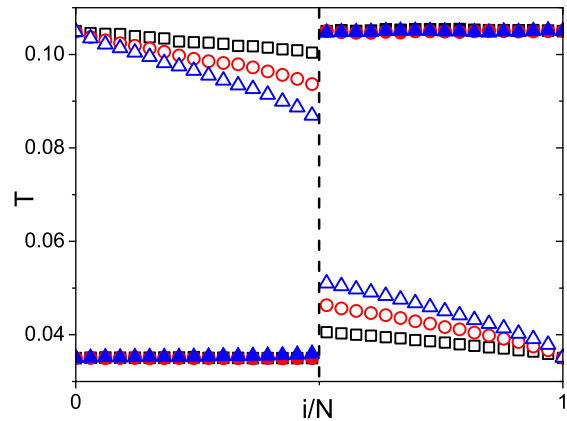}
\caption{\label{fig:jkint}(color online). Temperature profiles within a FK-FK thermal diode.
The temperature profiles for various interface elastic constant $k_{\text{int}}=0.01$ (square), $0.05$ (circle) and $0.2$ (triangle).
$T_0 = 0.07$, $N = 100$.
The open symbols correspond to a positive temperature bias $\Delta = 0.5$,
while the filled symbols correspond to negative temperature bias $\Delta = -0.5$.
The dashed line indicates the interface. The temperature jumps at the interface are due to the Kapitza resistance which
is addressed in  Sec.II.A.2.
Adapted from \textcite{li2004PRL93}.}
\end{figure}

For a given setup, the heat current through the system is mainly controlled by its
interface coupling strength $k_{\text{int}}$.
Figure~\ref{fig:jkint} depicts the temperature profiles for different $k_{\text{int}}$ and for two oppositely chosen bias strengths.
There occurs a large temperature jump at the interface. The size of the jump is larger for negative bias $\Delta$ (filled symbols in Fig.~\ref{fig:jkint}) than for positive bias $\Delta$. In the case  with negative bias   the temperature gradient inside each lattice segment  almost vanishes; implying that the resulting heat current is very small. This behavior is opposite to the case with  positive bias (open symbols in Fig.~\ref{fig:jkint}).

\subsubsection{Asymmetric Kapitza resistance}

The interface thermal resistance (ITR), also known as the Kapitza resistance,
measures the interfacial resistance to heat flow \cite{POLLACK1969RMP41,Swartz1989RMP61}.
It is defined as, $R\equiv{\Delta T}/{J}$,
where $J$ is the heat flow (per area) and $\Delta T$ denotes the temperature jump
between two sides of the interface.
The origin of the Kapitza resistance can be traced back to the heterogeneous electronic and/or vibrational properties of the different materials making up the interface at which the energy carriers become scattered.
The amount of relative transmission depends on the available energy states on either of the  two sides of the interface.
Originally, this  phenomenon was discovered by Kapitza in  1941 in  experiments detecting super-fluidity of He II \cite{Kapitza1941JP4}.

The high thermal rectification  in the above discussed model setups is mainly due to this interface effect.
In order to further improve the performance
\textcite{li2005PRL95} studied the ITR in a lattice consisting
of two weakly coupled, dissimilar anharmonic segments, exemplified by
a (FK)-chain segment and a neighboring Fermi-Pasta-Ulam (FPU)-chain segment.

Not surprisingly, the ITR in such a setup depends on the direction of the applied temperature bias.
A quantity that measures the degree of overlap of power spectra between left(L) and right(R) segments reads

\begin{equation}
S = \frac{\int_0^{\infty}
P_L(\omega)P_R(\omega)d\omega}{\int_0^{\infty}P_L(\omega)d\omega
\int_0^{\infty}P_R(\omega)d\omega}\;. \label{eq:Spm}
\end{equation}
Extensive  numerical simulations then reveal that $R_-/R_+ \sim \left(S_+/S_-\right)^{\delta_R}$ with
$\delta_R=1.68\pm 0.08$, and $|J_+/J_-| \sim
\left(S_+/S_-\right)^{\delta_J}$, with $\delta_J=1.62\pm 0.10$.
The notation $+/-$  indicates the case with a positive thermal bias, $\Delta>0$, and a negative thermal bias, $\Delta<0$, respectively.
These findings  thus support the strong dependence of thermal resistance on this very overlap of power spectra.

The physical mechanism of the asymmetric ITR between
dissimilar anharmonic lattices can therefore  be related to the match/mismatch of the corresponding power spectra.
As temperature  increases, the power spectrum of the FK lattice shifts downwards, towards lower frequencies.
In contrast, the power spectrum of the FPU-segment, however, shifts  upwards, towards higher frequencies. Due to these opposing  shifts, it  becomes evident that upon a reversing the thermal bias
the amount of match/mismatch  in such a FK-FPU setup can even exceed  that
of the above considered FK-FK setup. Conceptually this results in an even stronger thermal rectification.

\subsection{Negative differential thermal resistance:  The thermal transistor}

The design and the experimental realization of the thermal diode presents a striking first step towards the goal of doing phononics.
A next  challenge to be overcome is then a design  for a thermal transistor, which
allows for an {\it a priori} control of  heat flow much alike the familiar control of charge flow in a Field Effect Transistor.
Like its electronic counterpart, a thermal transistor
consists of three terminals: The drain (D), the source (S), and the gate (G).
When a constant temperature bias is applied between the drain and the
source, the thermal current flowing between source and drain
can be fine-tuned by the temperature that is applied to the
gate. Most importantly,  because the transistor is able to amplify a signal,
the changes in the heat current through the gate can induce an even larger current change from the drain to the source.

\begin{figure}[htbp]
\includegraphics[width=\columnwidth]{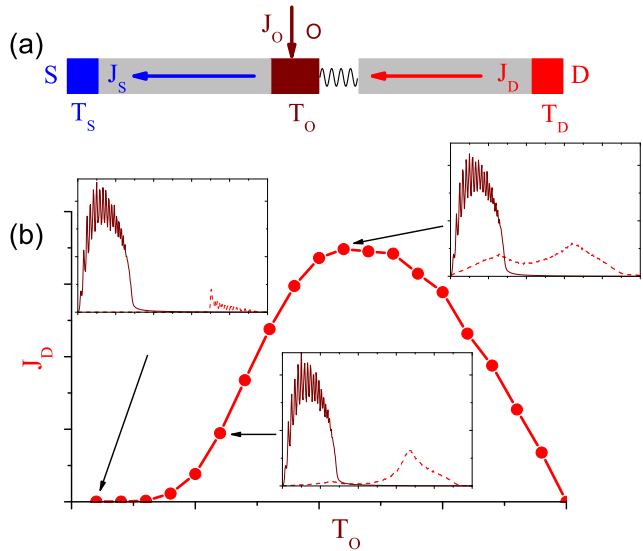}
\vspace{-0.8cm} \caption{\label{fig:NDTRscheme}(color online). Concept of a thermal transistor.
Panel (a). A one-dimensional lattice  is coupled at its two ends to  heat baths at temperature $T_{\text D}$ and $T_{\text S}$
with $T_{\text D} > T_{\text S}$.
A third heat bath with  temperature $T_{\text S} < T_{\text O} < T_{\text D}$ can be used to control the temperature at the node O, so as to control the heat currents $J_{\text D}$ and $J_{\text S}$.
Panel (b). In an extended  regime, as the temperature  $T_{\text O}$ increases, the thermal bias $(T_{\text D} - T_{\text O})$ at the interface decreases
while the power spectrum in the left segment of the control node (wine-color) increasingly matches with the power spectrum  of the neighboring segment that is connected to the drain D (red). This behavior is depicted with the insets where the corresponding power spectra of the left-sided and right-sided interface particles are depicted  for three representative values of $T_{\text O}$, as shown by the three arrows. The resulting drain current $J_{\text D}$ increases with increasing overlap until it reaches a maximum, and starts to decrease again.}
\end{figure}

Towards the eventual realization of such a thermal transistor device  we  consider the  concept depicted with  Fig.~\ref{fig:NDTRscheme}(a).  It uses a one-dimensional anharmonic lattice structure where the temperature at its ends
are fixed at $T_\text D$ and $T_\text S$ ($T_\text D>T_\text S$), respectively.
An additional, third heat bath at temperature $T_\text O$ controls the temperature at node O, so as to control the two heat currents $J_\text D$ and $J_\text S = J_{\text D} + J_\text O$. Let us next define the  current
amplification factor $\alpha$.  This quantity  describes the amplification ability of the thermal transistor,
as the change of the heat current
 $J_\text D$ (or $J_\text S$, respectively),  divided by the change in gate current $J_\text O$, which
acts as the control signal. This amplification quantity explicitly reads
\begin{equation} \label{amplification_factor}
\alpha = \left|{\partial J_\text D}/{\partial J_\text O}\right|.
\end{equation}
Eq.~(\ref{amplification_factor}) can be recast in terms of the differential thermal resistance of
the segment $S$, i.e.,
\begin{equation}
r_\text S \equiv \left({\partial J_\text S}/{\partial T_\text O}\right)^{-1}_{T_\text S = \text{const}}
\end{equation} and that of the neighboring segment $D$, i.e.,
\begin{equation}
r_\text D\equiv -\left({\partial J_\text D}/{\partial T_\text O}\right)^{-1}_{T_\text D = \text{const}}
\end{equation}
to yield
\begin{equation}
\alpha=\left|{r_\text S}/{(r_{\text S} +r_{\text D})}\right|.
\end{equation}
It can readily be deduced that if $r_\text S$ and $r_\text D$ are both positive then $\alpha$ emerges to be less than unity.
Consequently, such a thermal transistor cannot  work.

For a transistor to work it is thus necessary that the current amplification factor obeys  $\alpha>1$.  This implies a {\it negative differential thermal resistance} (NDTR); i.e.,  it requires a transport regime  wherein the heat current {\it decreases} with {\it increasing} thermal bias. Such behavior occurs with the thermal diode characteristics depicted with Fig.~\ref{fig:jdelta} (d); note the behavior for the case with filled (blue) triangles with $\Delta$ increasing from $\sim [-0.5, - 0.2]$. It should be pointed out that such an (NDTR)-behavior is in no conflict with any physical laws \footnote{Note that this NDTR should not be confused with  absolute negative thermal resistance around an equilibrium working point at thermal bias $\Delta=0$. The latter would imply that heat flows from cold to hot which violates the principle of Le Chatelier-Brown \cite{Callen1960book}, i.e. no {\it opposite} response to a small external perturbation around a thermal equilibrium is possible. Such an anomalous response behavior is, however,  feasible when the system is taken (at a zero bias) into a stationary non-equilibrium state \cite{hanggi2009RMP81}.}.

While negative differential electric resistance has long been realized  and extensively studied  \cite{Esaki1958PR109},
the concept of NDTR has been  proposed more recently only \cite{li2004PRL93,li2006APL88}.
With temperature $T_O$ increasing, the sensitively temperature-dependent power spectra of the two segments match increasingly better which not only offsets the effect of a decreasing thermal bias $(T_D-T_O)$ but even induces an increasing heat current.
This behavior is illustrated in Fig.~\ref{fig:NDTRscheme}(b).

\begin{figure}[!htbp]
\includegraphics[width=\columnwidth]{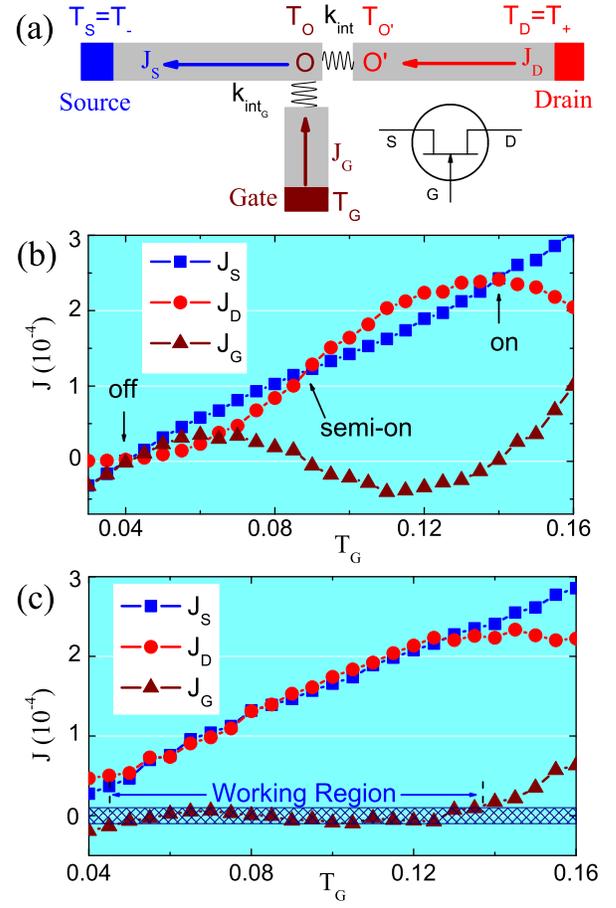}
\caption{\label{fig:switch_modulator} (color online). Thermal transistor.
(a) Sketch of a  thermal transistor device.
Just as in the case of an electronic transistor, it consists of two
segments (the Source and the Drain) and, as well,  a third segment (the
Gate) where the control signal is injected.
The temperatures $T_{\text D}$ and $T_{\text S}$ are fixed at high, $T_+$, and low, $T_-$, values.
The negative differential thermal resistance (NDTR) occurs at the interface between O and O'. This part then  makes it possible that over a wide regime of parameters,
when the gate temperature $T_{\text G}$ rises, not only $J_{\text S}$ but also $J_{\text D}$ may increase. Using different system parameters allow then
different function, this being either a  thermal switch (panel (b)) or also a thermal  modulator (panel (c)).
Panel (b). Function of a thermal switch: At three working points where $T_{\text G}$ = $0.04$, $0.09$ and $0.14$,
the heat current $J_{\text D}$ equals $J_{\text S}$, yielding  $J_{\text G} = J_{\text S} - J_{\text D}$ vanishing identically.
These three working points correspond to (stable) ``off'', (unstable) ``semi-on'' and (stable) ``on'' states, at which
$J_{\text D}$ differs substantially.
We can switch, i.e., forbid or allow heat  flowing  by setting $T_{\text G}$ at these different values.
Panel (c). Function of a thermal modulator. Over a wide temperature interval of gate temperature
$T_{\text G}$, depicted via the hatched working region, the heat current $J_{\text G}$ remains very small, i.e. it remains inside the hatched regime,
while the two heat currents $J_{\text S}$ and $J_{\text D}$ can be
continuously controlled from low to high values.
Adapted from \textcite{li2006APL88}.
}
\end{figure}

A system displaying NDTR constitutes the main ingredient for  operation of a  thermal transistor.
The scheme of a thermal transistor is shown in Fig.~\ref{fig:switch_modulator}(a).
In order to make this setup physically  more realistic, a third segment (G) is  connected to the node O. This is done so, because in an actual device it is difficult to directly control the temperature of node O,  being located  inside the device.
Using different sets of parameters, this  thermal transistor  can
work either as a thermal switch, Fig.~\ref{fig:switch_modulator}(b), or also as a thermal modulator, Fig.~\ref{fig:switch_modulator}(c).

The key prerequisite for a thermal transistor, i.e. the NDTR phenomenon has been investigated in various other systems recently,
e.g., for high dimensional lattice models \cite{Lo2008JPSJ77}. A gas-liquid transition has also been utilized
in the design of a thermal transistor \cite{Komatsu2011PRE83}.
The condition for the existence of a NDTR regime is more stringent than that of obtaining merely thermal rectification.
The crossover from existence to nonexistence of NDTR have been  investigated for a set of different  lattice structures,
both by analytical and numerical means  \cite{shao2009PRE79,He2009PRB80}.
Because the NDTR in these lattice models basically derives  from an interface effect, it is plausible that with a too large
interface coupling strength $k_{\text{int}}$  or a too long  lattice length,
it is rather the thermal resistance of the involved segments than  the interface resistance that rules the NDTR. As a consequence, the NDTR effect typically becomes  considerably  suppressed.
We therefore expect that NDTR can  experimentally be realized with  nano-scale materials, for example using  nano-tubes, nano-wires, etc., see \cite{Chang2006Sci314,Chang2007PRL99,chang2008PRL101}. This issue is addressed in greater detail with Sect. III below.

\subsection{Thermal logic gates}

The phenomenon of NDTR provides not only the function for thermal switching and
thermal modulating, but also the
essential input  in devising thermal logic gates.
Setups for all major thermal logic gates able to perform logic operations
have been put forward recently \cite{wang2007PRL99}.

In  digital electric circuits, two boolean states ``1'' and ``0'' are
encoded by two different values of voltage, while in a
digital thermal circuit these boolean states can be defined by two different values of temperature $T_{\text{on}}$ and $T_{\text{off}}$. In the following we discuss the way to realize such  individual logic gates.

A most fundamental logic gate is the signal repeater which
`digitizes' the input. Namely, when the input temperature is lower/higher than a critical value $T_c$, with
$T_{\text{off}} < T_c < T_{\text{on}}$, the output is set at  $T_{\text{off}}$/$T_{\text{on}}$, respectively.
This does not present a simple task;  it must take into account that small errors may accumulate,
thus leading eventually to  incorrect outputs.

The thermal repeater can be obtained  by use of thermal switches.
Let us inspect again the   thermal switch shown in Fig.~\ref{fig:switch_modulator}(a),
in which the $T_{\text G}$-dependence of the heat currents $J_{\text D}$, $J_{\text S}$ and $J_{\text G}$ is illustrated with Fig.~\ref{fig:switch_modulator}(b).
When the gate temperature $T_{\text G}$ is set very close but not
precisely at $T_{\text{off}}$/$T_{\text{on}}$, then
the heat current in the gate segment makes the temperature
in the junction node O approach more closely  $T_{\text{off}}$/$T_{\text{on}}$.
Therefore, upon connecting
such switches in series, which involves plugging the
output (from node O) of one switch into the gate of the next one,
the final output will  asymptotically approach that of an ideal repeater,
i.e. $T_{\text{off}}$/$T_{\text{on}}$, whichever is closer to the input temperature $T_\mathrm{G}$.

\begin{figure}
\includegraphics[width=\columnwidth]{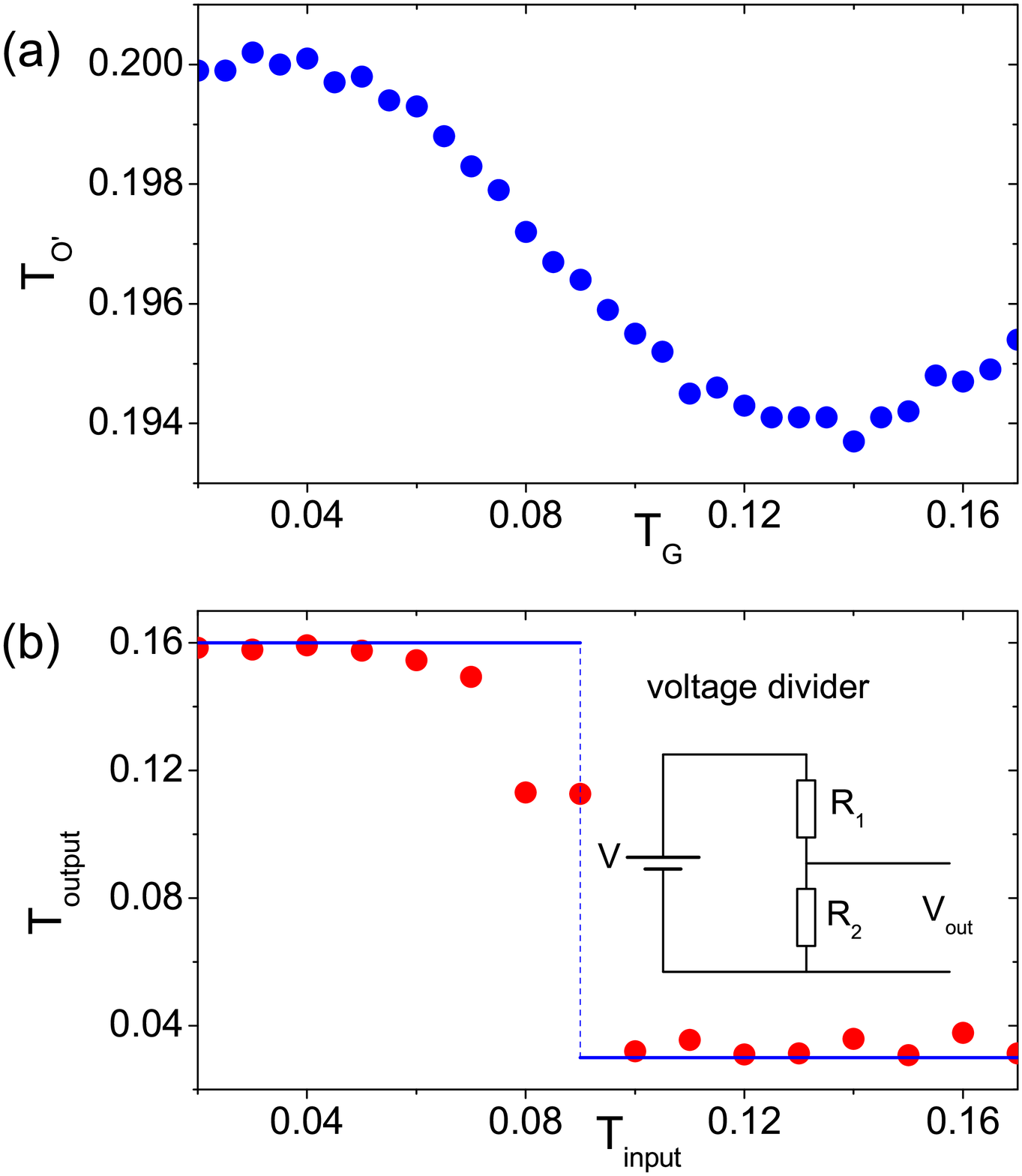}
\vspace{-0.5cm} \caption{\label{fig:negresp_notgate} (color online). Negative response and thermal NOT gate.
(a) Temperature of the node O' as a function of $T_\mathrm{G}$ for the setup shown in Fig.~\ref{fig:switch_modulator}(a).
In a wide range of $T_\mathrm{G}$, $T_{\mathrm{O'}}$ decreases as $T_G$ increases.
(b) Function of the thermal NOT gate. The thin (blue) line indicates the function of an ideal NOT gate.
Inset: Structure of a two-resistor voltage divider, the
counterpart of a temperature divider, which supplies a voltage
lower than that of the battery. The output of the
voltage divider is: $V_{\textup {out}}=VR_2/(R_1+R_2)$.
Adapted from \textcite{wang2007PRL99}.}
\end{figure}

A NOT gate reverses the input; it yields  the response ``1'' whenever it  receives ``0'',
and vice versa. This requires that the
output temperature falls when the input temperature
rises, and vice versa.
This feature can be realized by injecting the signal from the node G
and collecting the output from the node O', cf. Fig.~\ref{fig:switch_modulator}(a).
The NDTR between the nodes O and O' again plays the key role.
A higher temperature $T_{\text G}$ induces a larger thermal current
$J_{\text D}$ and therefore increases the temperature bias in segment D.
$T_{\text {O'}}$ thus decreases (notice that $T_{\text D}$ is fixed) and a
negative response is thus realized, cf. Fig.~\ref{fig:negresp_notgate}(a).
Suppose $T_{\text {O'}}$ equals $T_{\text {O'}}^{\text{off}}$ and $T_{\text {O'}}^{\text {on}}$, with $T_{\text {O'}}^{\text{off}}$$ > $$T_{\text {O'}}^{\text{on}}$,
when $T_{\text G}$ equals $T_{\text{off}}$ and $T_{\text{on}}$, respectively.
A remaining problem is that both $T_{\text{O'}}^{\text{off}}$ and $T_{\text {O'}}^{\text{on}}$ are
higher than $T_c$ (in fact even higher than $T_{\text{on}}$). This in turn  will be
always treated as a logical ``1'' by the following device.
This problem can be solved if we apply a ``temperature divider'', the  counterpart of a voltage
divider, which is depicted in inset of
Fig.~\ref{fig:negresp_notgate} (b).
Its output equals a ratio of
its input. By adjusting this ratio,
we can make the output of the temperature divider be
higher/lower than $T_c$ when its input equals
$T_{\text {O'}}^{\text{off}}$/$T_{\text {O'}}^{\text{on}}$.
We then  employ a thermal repeater to digitize the output from the temperature divider.
In doing so the function of a NOT gate is therefore realized, as
depicted with  Fig.~\ref{fig:negresp_notgate} (b).

Similarly, an AND gate is a three-terminal (two inputs and one output) device.
The output is ``0'' if either of the inputs are ``0''.
Because we now have the thermal signal repeater at our disposal, an AND gate is readily realized by
plugging two inputs
into the same repeater. It is clear that when both inputs are
``1'', then the output is also ``1''; and when both inputs
are ``0'', then the output is also ``0''. By adjusting
some parameters of the repeater, we are able to make the final output to be ``0'' when the two
inputs are opposite, therefore realizing a thermal AND gate.
An OR gate, which exports ``1'' whenever the two
inputs are opposite, can also be realized in the similar way.

\subsection{Thermal memory}

Towards the goal of an all phononic computing, yet another indispensable element, besides thermal logic
gates, are thermal memory elements that enable the storage of information via its encoding by heat or temperature.

A possible such setup acting as a thermal memory  has been proposed by
\textcite{wang2008PRL101}. Its blueprint has much in common with the working scheme for the
thermal transistor, cf. Fig.~\ref{fig:switch_modulator}(a).
Adjusting parameters appropriately, negative differential thermal resistance
can be induced between the chain segments connecting to node O and node O', respectively.
With fixed $T_{\text S}$ and $T_{\text D}$, obeying $T_{\text S} < T_{\text D}$, and a heat bath at temperature $T_{\text G}$ that is coupled to the node G, three possible  working points  for $T_{\text G}$, i.e, $T_{\text{off}}$, $T_{\text{semi-on}}$ and $T_{\text{on}}$ can be realized. At these three working points the gate current vanishes, $J_{\text G} = 0$, thus balancing perfectly $J_{\text S}$ and
$J_{\text D}$. Upon analyzing the slope of $J_{\text G}$ at these very points two of these, i.e., ($T_{\text{off}}, T_{\text {on}}$) denote stable working points and the intermediate one, $T_{\text {semi-on}}$, is unstable.  Notably, these working states remain stationary when the heat bath that is coupled to terminal G is removed. Now, however, the corresponding temperatures at these working points exhibit fluctuations. As is well known from stochastic bistability \cite{Hanggi1990RMP62}, small fluctuations around these working points drive the system consistently back to the stable working points and away from the unstable working point. Accordingly, the system possesses two long lived meta-stable states, $T_{\text O} = T_{\text{on}}$ and $T_{\text O}$=$T_{\text{off}}$, while $T_{\text O} = T_{\text{semi-on}}$ is unstable.


The stability of these so adjusted  thermal states at $T_{\text {on}}$, $T_{\text {off}}$ implies that these states remain basically unchanged over an extended time span  in spite of  thermal fluctuations. This holds true even if a small external perturbation, as for example imposed by a small thermometer, reading the temperature  at the node O, is applied.

\begin{figure}[ht]
\includegraphics[width=\columnwidth]{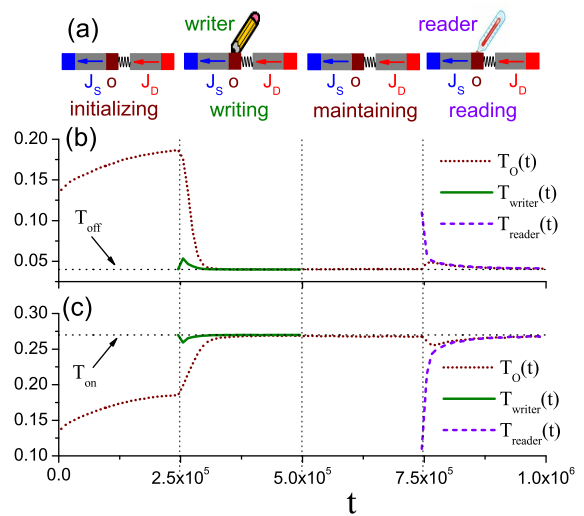}
\vspace{-0.5cm} \caption{\label{fig:memory_writing_reading} (color). Write-and-Read cycle in a thermal memory.
(a) Illustration of the four stages of a complete writing-reading process.
The left and right ends of this thermal memory are coupled to heat baths at fixed temperatures $T = 0.03$ and $T = 0.3$, respectively.
During the initializing  stage the system relaxes towards its stationary state with the temperature at site O being $T_{\text O} = 0.18$. The different stages during this cycle are indicated by separating dashed vertical lines in panels (b) and (c).
(b) Time evolution of the local temperature $T_{\text {O}}(t)$ (dotted wine color) of the central particle at site O,  the temperature of the Writer $T_{\text {writer}}(t)$ (solid olive)  and the reader $T_{\text {reader}}(t)$ (dashed violet) during a complete Write-and-Read  process.
The  time-dependent local temperature of the particle (with unit mass) at site O is obtained  by integrating and averaging the squared velocity $v_{\text O}^2(t)$  over a short time span ($10^4$ time steps), centered  at time-instant t.
Writer and Reader  are realized with  short lattice segments, while $T_{\text{writer}}$ and $T_{\text{reader}}$ are the averaged temperatures of all  particles within the Writer and Reader, respectively.
During the writing stage the node O is coupled to one end of the Writer whose opposite end is
coupled to a heat bath set at the ``off''-value $T_{\text{off}} = 0.04$.
During the reading stage, the reader set at initial temperature $0.11$ is attached to the node O.
(c) The opposite Write-and-Read cycle with the Writer set at the ``on''-value with
$T_{\text{on}} = 0.27$. Adapted from \textcite{wang2008PRL101}.}
\end{figure}

The working principle of a Write-and-Read cycle of this so designed  thermal memory is depicted with Fig.~\ref{fig:memory_writing_reading}:
Starting out at time  $t=0$ from a random initial preparation of all particles making up the memory device  the local temperature at site O  relaxes to its stationary value  $T_{\text O} \sim 0.18$ (initializing stage).  For the case in panel (b) in Fig.~\ref{fig:memory_writing_reading} the Writer, being prepared at  its boolean temperature value $T_{\text{off}}$  is next connected to the site O (writing stage). As  can be deduced from panel Fig.~\ref{fig:memory_writing_reading}(b),  the temperature $T_{\text O}$ quickly relaxes during this writing cycle to this boolean value, and maintains  this value over an extended time  span, even  after the Writer is removed (maintaining stage).
More importantly, $T_{\text O}$  self-recovers to this
setting temperature $T_{\text{off}}$ after being
exposed  to the small perturbation as induced by the Reader (a small thermometer) during the following
data-reading stage. The data stored in the thermal memory are
therefore precisely read out without destroying the memory state.

In panel  Fig.~\ref{fig:memory_writing_reading}(c), the Write-and-Read
process corresponding to the opposite boolean writing temperature value, i.e. $T_{\text O}= T_{\text {on}}$, is depicted.
Accordingly, this so engineered  Write-and-Read cycle with its two possible writing
temperatures $T_{\text {off}}$ and $T_{\text {on}}$  does accomplish
the task  of a thermal memory device.

\section{Putting Phonons to Work } \label{Sec III}

In the foregoing Sec. II, we have investigated various setups involving stylized nonlinear lattice structures in order to manipulate heat flow. We next discuss how to put these concepts into practical use with physically realistic systems. In doing so we consider numerical studies of suitably designed  nanostructures which exhibit the designated thermal rectification properties. This discussion is then followed with first experimental realizations of a thermal diode and the thermal memory.

Among the plenty of physical materials that come to mind,  low dimensional nanostructures like nanotubes,
nanowires  and graphene likely offer  optimal choices to realize the desired thermal rectification features obtained with nonlinear lattice studies. It is known from  theoretical studies \cite{Lepri2003PR377, Dhar2008AdvPhys57,Saito2010PRL104} and  experimental validation \cite{chang2008PRL101,ghosh2010NatureMaterials9}, that the characteristics of heat flow (such as the validity of the Fourier Law \cite{Lepri2003PR377, Dhar2008AdvPhys57,Saito2010PRL104}) sensitively depends on  spatial dimensionality and the absence or presence of (momentum) conservation laws.

This is even more the case for nano materials, wherein due to the limited size, the discrete phonon
spectrum \cite{yang2010NT5} results in a distinct dependence of the thermal quantities on the specific geometrical configuration, mass distribution and ambient temperature. Overall, this makes nanosized materials promising candidates for phononics. Because the experimental  determination  of thermal transport properties is not straightforward the combined use of theory and numerical simulation is indispensable in devising phononics devices. Moreover, novel atomistic computational algorithms have been developed which facilitate the study of experimentally relevant system sizes \cite{Li2010PRB82}.

\subsection{Thermal diodes from asymmetric nanostructures}

\begin{figure}[htb]
\includegraphics[width=\columnwidth]{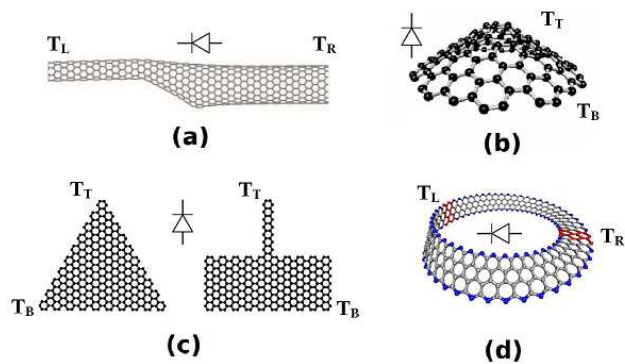}
\caption{\label{fig:Fig4-1} (color online). Schematic system setups for  thermal diodes devised from
asymmetric nano structures:
(a) a typical (n, 0)/(2n, 0) carbon nanotube junction; (b) a carbon-nanocone based thermal rectifier;
 (c) two asymmetric graphene nanoribbons (GNRs), a trapezia-shaped GNR and, as well, a structure made of two-rectangular GNRs  of different width; (d) a topological configuration based on a  M\"obius graphene stripe. The red parts denote the corresponding heat bath regions. The bath temperature of two ends are
denoted as $T_{T}$ (Top) and $T_{B}$ (Bottom) in (b) and (c), and  as $T_{L}$ (Left) and $T_{R}$ (Right) in (a) and (d), respectively. These figures are adapted from \textcite{wu2007PRB76}, \textcite{Yang2008APL93}, \textcite{Yang2009APL95}, \textcite{Jiang2010EPL89}.}
\end{figure}

Carbon nanotubes (CNTs) have recently attracted  attention  for  applications in nanoscale electronics, mechanical and thermal devices. CNTs posses a high  thermal conductivity at room temperature \cite{kim2001PRL87} and  phonon mean free paths that extend over the length scale of structural ripples; thus providing  ideal phonon waveguide properties \cite{Chang2007PRL99}.

Let us consider  thermal rectification in
single-walled carbon nanotube (SWCNT) based
junctions \cite{wu2007PRB76}. A typical (n,
0)/(2n, 0) intramolecular junction structure is shown
in Fig. \ref{fig:Fig4-1}(a) in which the structure contains two
parts, namely, a segment with a (n, 0) SWCNT and a segment made of  a (2n, 0)
SWCNT. These two segments are connected by \emph{m} pairs of
pentagon-heptagon defects. By use of non-equilibrium molecular
dynamics (NEMD) simulations, one finds that the heat flux  from the (2n, 0)
to (n, 0) tube exceeds that from the (n, 0) tube to the (2n, 0) segment.
The corresponding thermal rectification increases upon raising the
temperature bias.

Another beneficial feature is that the
rectification is only weakly dependent on the detailed structure of the
interface, assuming that the connecting region is sufficiently short. This is so
because heat is predominantly carried by   long
wavelength phonons, which are scattered mainly by large defects. Just as with nonlinear lattice models, the match/mismatch
of the energy spectra around the interface is the underlying
mechanism for rectification. Furthermore, in the elongated
structure \cite{Wu2008JPCM20} the heat flux becomes smaller than
that of the non-deformed structure; its temperature
dependence, however, becomes more pronounced, due to the intrinsic tensile stress.

The carbon nanocone depicted in Fig. \ref{fig:Fig4-1}(b) is yet another carbon based
material exhibiting a high asymmetric geometry. Its thermal properties were
investigated by \textcite{Yang2008APL93}. A temperature difference, $\Delta$, between the two ends of
nanocone is introduced, being positive when the bottom of nanocone is at a higher
temperature, and is negative in the opposite case. It was found that the nanocone behaves as a
``good'' thermal conductor under positive thermal bias and as a ``poor''
thermal conductor when exposed to a negative thermal bias. This suggests
that the heat flux runs preferentially along the direction of
decreasing diameter.

In order to compare the impact of mass-asymmetry versus  geometry-asymmetry for
thermal rectification a  nanocone structure with a graded
mass distribution was discussed \cite{Yang2008APL93}. The mass of the atoms of the
nanocone are devised to change linearly; that is, the top atoms possess a minimum mass
$M_{C12}$ and the bottom atoms are at the maximum mass of $4M_{C12}$, with
$M_{C12}$ being the mass of $^{12}C$ atom. NEMD results then yield a
rectification ratio,   $|J_+ - J_-|/|J_-|$, that is $10{\%}$ for a nanocone with uniform masses and is
$12{\%}$ for the mass-graded  nanocone (at identical thermal bias $\Delta$).
This  2{\%} increase for the mass-graded distribution evidences that the role of
geometric asymmetry is more effective in boosting thermal rectification.

These geometric asymmetric SWCNT junctions all exhibit  thermal
rectification. Compared to the SWCNT setup, it is easier, however, to
control the shape of graphene by the nano cutting technology such as helium ion microscope. Graphene nanoribbons
(GNRs), cf. Fig. \ref{fig:Fig4-1}(c), present promising elements for  nanoelectronics and nanophononics.
For instance, \textcite{Yang2009APL95} have demonstrated  tunable thermal conduction in GNRs. They studied the
direction dependent heat flux in asymmetric structural GNRs, and
explored  the impacts of both, GNR shape and size,  for the rectification
ratio. Two types of GNRs were considered: trapezia-shaped GNR and two-rectangular GNRs of different width, Fig. \ref{fig:Fig4-1}(c).

These two types of GNRs behave as a ``good'' thermal
conductors under positive thermal bias (i.e. bottom end at higher temperature) and as a ``poor'' thermal
conductor under negative thermal bias (i.e. top end at higher temperature). This finding  is similar to the
rectification phenomena observed in carbon nanocone structures
\cite{Yang2008APL93}, which again results from the amount of  match/mismatch between
the respective phonon spectra.

It is interesting to note that the rectification
ratios of GNRs are substantially larger than those in nanocone and SWCNT
junctions. With $\Delta$=0.5, the rectification ratio of the nanocone is
$96{\%}$ \cite{Yang2008APL93} and that of SWCNT intramolecular
junction is only around $15{\%}$\/ \cite{wu2007PRB76}, while the
rectification ratio in GNRs  is about $270{\%}$ and $350{\%}$ for two-rectangular GNRs
and the trapezia-shaped GNR, respectively. The rectification ratio of
trapezia-shaped GNR is thus larger than that of two-rectangular GNR under the same thermal bias
difference. This feature is consistent with the phenomenon that the carbon
nanocone (i.e. the geometric graded structure) possesses a larger rectification ratio
than the two-segment carbon nanotube (n, 0)/(2n, 0) intra-molecular junction.
The phonon spectra change continuously in the graded structures which implies a more efficient control
of heat flux. Similar rectification features were  observed
in graphene nanoribbons by \textcite{Hu2009NL9}.

Yet another advantage of a GNR based thermal rectifier is its weak dependence on length. Because energy is transported
ballistically in graphene, the heat flux is essentially independent on
size. Both $J_{+}$ and $J_{-}$ are saturated when the GNR length is
longer than $\sim$ 5.1 nm, yielding that the rectification ratio remains
constant around  92{\%}. Moreover, compared to a setup composed of a single-layer
graphene, an even larger rectification ratio can be achieved in
few-layer asymmetric structures, this being a consequence of layer-layer interactions \cite{Zhang2011NS3}.

Thermal rectification can also be realized
with a topological graphene setup such as the M\"obius graphene strip
\cite{Jiang2010EPL89}, Fig. \ref{fig:Fig4-1}(d). The advantage of this topology induced
thermal rectification is that it is practically insensitive to  temperature and size of
the system. In this structure, the asymmetry originates from the intrinsic
topological configuration.

Because  phonons are strongly
scattered due to the mismatch in vibrational properties of the
materials forming the interface, which in addition depends on the sign of applied thermal bias,
this produces  an asymmetric Kapitza contact resistance. Capitalizing on this idea
a silicon-amorphous polyethylene thermal rectifier was designed
\cite{Hu2008APL92}. The result is that the heat current from the polymer to the
silicon is larger than vice versa. To examine the origin of the
thermal rectification, the density of states of phonons on each side
of the interface was investigated. The phonon density of
states  significantly  softens in the ploymer as it becomes warmer.
This increases the density of states in the polymer at low frequencies. Those
low frequency acoustic modes in silicon thus increase their transmission probability, yielding an
enhanced thermal transport.

The setups discussed thus far  are all operating in steady state.
A transient thermal transport exhibiting a time-dependent   thermal rectification has been investigated for a
Y-SWCNT junction by \textcite{Noya2009PRB79}. From their molecular dynamics simulation it is
reported that a heat pulse propagates unimpeded from the stem to the
branches, undergoing little reflection. For  the reverse temperature
bias, however, there is a substantial reflection back into the branches.
The transmission coefficient from the stem to the  branches is more than four times that of
the reverse direction.

Let us reflect again on the choice of materials suitable in obtaining those theoretically predicted thermal rectification features. A thermal rectifier constitutes  a two-terminal thermal device whose working principle rests upon  the different overlap of the temperature dependent phonon spectra. For this to occur strong anharmonicity plays an important factor. Given the above discussion of geometric asymmetric structures a strong intrinsic anharmonicity corresponds to diffusive phonon transport. However, diffusive phonon transport presents not a necessary condition for thermal rectification. In the asymmetric GNRs, phonon transport is almost  ballistic. 
The conclusion is that thermal rectification exists for  both, ballistic and diffusive phonon transport, assuming that a match/mismatch of the phonon spectrum of the two ends is present. This difference in phonon spectra can originate from different sources, as for example  from a asymmetric mass distribution, different geometry, size or spatial dimension.

Despite an abundance of parameter dependent rectification studies \cite{Wu2009PRE80, Wu2009PRL102, Yang2007PRB76, Hu2006PRE74}, these were rarely tested against experiments. All those NEMD-studies rely on  mathematically idealized material Hamiltonians, which in practice may still be far from physical reality.  Only {\it in situ} experiments thus  present the ultimate test bed for a validation of the wealth of available theoretical results and, even more importantly, dictate the necessary next steps towards an implementation of phononics.

\subsection{{\it In situ} thermal diodes from mass-graded nanotubes: Experiment}

\begin{figure}[htb]
\includegraphics[width=6cm]{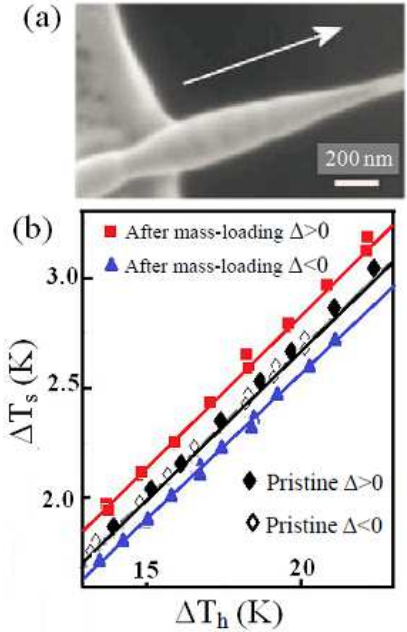}
\vspace{-0.2cm} \caption{\label{fig:Fig4-4}(color online). Experimental realization of
thermal rectification in nanotube junctions. (a) Scanning-Electron-Microscope images of boron nitride nanotubes (BNNTs) after deposition of $C_9$$H_{16}$$P{\text t}$. (b) Graphical representation of the temperature changes of the heater ($\Delta T_{h}$) and sensor ($\Delta T_{s}$) for the nanotubes before and after deposition of $C_9$$H_{16}$$P{\text t}$. Adapted from \textcite{Chang2006Sci314}.}
\end{figure}

On the experimental side, a pioneer thermal diode work
has been performed by \textcite{Chang2006Sci314}. In their
experiment, cf. Fig. \ref{fig:Fig4-4}, carbon nanotubes (CNTs) and boron nitride
nanotubes (BNNTs) were gradually deposited on the surface with the heavy
molecules located along the length of the nanotube in order to establish an asymmetric
mass distribution. As an important test, it has been experimentally checked that in unmodified NTs
with uniform mass distribution the thermal conductance is indeed
{\it independent} of the direction of the applied thermal bias. In clear contrast, however,
the inhomogeneous NT-system in fact does exhibit  asymmetric axial thermal
conductance, with greater heat flow in the direction of decreasing
mass density. The thermal rectification ratios, i.e.  $|J_+-J_-|/|J_-|$,  were measured as $2{\%}$
and $7{\%}$ for CNT and BNNT, respectively. The higher thermal
rectification found in BNNT might originate from
stronger nonlinearity, as induced by the ionic nature of the B-N bonds, which in turn
favors  thermal rectification.

In order to understand the mechanism of rectification in Chang {\it et al's} experiment,  subsequent work \cite{Yang2007PRB76} has studied the thermal properties in a one-dimensional anharmonic lattice with a mass gradient. \textcite{Yang2007PRB76} found that in the 1D mass-graded chain, when the heavy-mass end is set at high temperature, the heat flux is larger than that under the reverse temperature bias. This is  consistent with Chang's experimental results \cite{Chang2006Sci314}. Also, the larger is the mass gradient the larger is thermal rectification. As mentioned before, this can be explained from differences in overlap of the respective  phonon power spectra.

%


\subsection{Solid-state-based thermal memory: Experiment}
\begin{figure}
\includegraphics[width=\columnwidth]{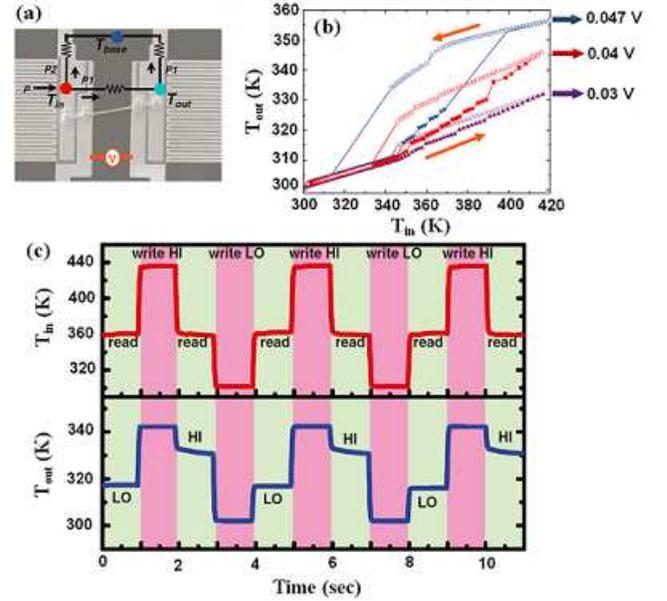}
\vspace{-0.2cm} \caption{\label{fig:Fig4-3}(color online). Experimental realization of the
thermal memory. (a) Scanning-Electron-Microscope image of the memory device, consisting of a $VO_{2}$ nanobeam connecting the input terminal (left side) and output terminal (right side). An equivalent thermal circuit is also depicted in the image. Panel (b). Output temperature, ${\textup T_{\textup {out}}}$, as a function of input temperature, ${\textup T_{{\textup {in}}}}$, upon heating (filled symbols) and cooling (open symbols). The  characteristics of the thermal memory is tuned under three different voltage biases, as indicated in the panel. Panel (c). Write and Read out process of  logical ``1'' (= HIGH temperature state) and ``0'' (= LOW temperature state).
Upper part, ${\textup T_{{\textup {in}}}}(t)$ for writing  (red line): To write a HIGH (=1) temperature state to the output terminal, a heating pulse generated by increasing the heating current was applied to the input terminal to set ${\textup T_{{\textup {in}}}}$ to $435$K. To switch the output terminal to a LOW  temperature state, a
cooling pulse was applied to the input terminal to set ${\textup T_{{\textup {in}}}}$ to $300$K
by natural cooling of ${\textup T_{{\textup {in}}}}$  to the substrate temperature ${\textup T_{{\textup {base}}}}$. Lower part, ${\textup T_{{\textup {out}}}}(t)$ for reading (blue line): When
${\textup T_{{\textup {in}}}}$ was returned to $360$K for reading, the HIGH (LOW) state of
$\sim ~ 332.1 {\pm 0.1}$ K (${\sim ~} 317 {\pm 0.8}$K) was retained and could be read out at
the output terminal ${\textup T_{{\textup {out}}}}$. Here the voltage bias is $0.04$V. Data and Figure adapted from \textcite{Xie2011AM21}.}
\end{figure}

Apart from the experimental demonstration of a nanoscale thermal rectifier, a thermal  memory device has
recently also brought into operation experimentally by \textcite{Xie2011AM21}.
In their work, see in Fig. \ref{fig:Fig4-3}(a), a
single-crystalline $VO_{2}$ nanobeam is used to store and retain
thermal information with two temperature states as input ${\textup T_{{\textup {in}}}}$
and output ${\textup T_{{\textup {out}}}}$, which serve as the logical boolean units ``1''(= HIGH) and ``0''(=LOW), for writing and reading. This has been achieved by
exploiting a metal-insulator transition. A nonlinear
hysteretic response in temperature was obtained in this way.
A  voltage bias across the nanobeam has been applied to tune the
characteristics of the thermal memory.  One finds that the
hysteresis loop becomes substantially enlarged, see in  Fig.
\ref{fig:Fig4-3}(b),  and is shifted towards lower temperatures with
increasing voltage bias. Moreover, the difference in the output
temperature ${\textup T_{{\textup {out}}}}$ between its ``HIGH'' and ``LOW'' temperature states increases substantially
with increasing voltage bias.
To demonstrate the repeatability of the thermal memory, they have performed repeated Write(High)-Read-Write(Low)-Read cycles using heating and cooling pulses to the input terminal, see panel (c) in Fig. \ref{fig:Fig4-3} for more details. Repeated cycling over 150 times proves the reliable and repeatable performance of this thermal memory.




\section{Shuttling Heat and Beyond} \label{Sec IV}

The function of the various thermal electronic analog devices discussed in Sect. II used a heat control mechanism which is based on a static thermal bias.
In order to obtain an even more flexible control of heat energy, being  comparable with the richness available known for electronics, one may design intriguing phononic devices which utilize  temporal, ac gating modulations as well. Such forcing makes possible the realization of a plenitude of  novel phenomena such as the heat ratchet effect, absolute negative heat conductance or the realization of Brownian (heat) motors, to name but a few \cite{astumian2002PhysicsToday55,hanggi2005AnnalsPhysics14,hanggi2009RMP81}. Among the necessary prerequisites to run such heat machinery are thermal noise, nonlinearity, unbiased nonequilibrium driving of deterministic or stochastic nature and a  symmetry
breaking mechanism. This then carries the setup away from thermal equilibrium, thereby
circumventing the second law of thermodynamics, which otherwise would impose a vanishing directed transport \cite{hanggi2009RMP81}.

Dwelling on similar ideas used in Brownian motors for directing particle flow, an efficient pumping or
shuttling of energy across spatially extended
nano-structures can be realized via modulating either one or more  thermal bath temperatures, or applying
external time-dependent fields, such as mechanical/electric/magnetic forces.
This gives rise to  intriguing phononic phenomena such
as {\it a priori} directed shuttling of heat {\it against} an external  thermal bias or the pumping of heat that is assisted by a geometrical-topological component.

\subsection{Classical heat shuttling}
In the following we shall elucidate the objective for shuttling heat against an externally applied thermal bias.
A salient requirement for the {\it modus operandi} of heat motors is
the presence of a spatial or dynamic symmetry breaking.

A possible scenario consists in coupling  an asymmetric nonlinear structure to two baths; i.e.,
a left(L) and right(R) heat bath which can be modeled by classical Langevin
dynamics. Applying a  periodically time-varying temperature in one or both heat baths,
$T_{\text{L(R)}}(t)=T_{\text{L(R)}}(t+2\pi/\omega)$, possessing the same
average temperature $\overline{T_{\text{L(R)}}(t)}=T_0$, then brings the system out-of-equilibrium.
Noteworthy is that this so driven system is unbiased; i.e., it exhibits a vanishing average thermal bias
$\overline{\Delta T(t)}=\overline{T_{\text{L}}(t)-T_{\text{R}}(t)}=0$.
The asymptotic
heat flux $J(t)$ assumes the periodicity of the external driving
period $2\pi/\omega$ and the time-averaged heat flux $J$ follows as the
cycle-average over a full temporal period:
$J=\frac{\omega}{2\pi}\int^{2\pi/\omega}_{0}J(t)dt$. Consequently, a  nonzero
average heat flux $J\neq 0$  emerges which then provides the seed to even shuttle
heat {\it against} a net  thermal bias $\overline{\Delta T(t)}\neq0$.

\begin{figure}
\centerline{\includegraphics[width=7cm]{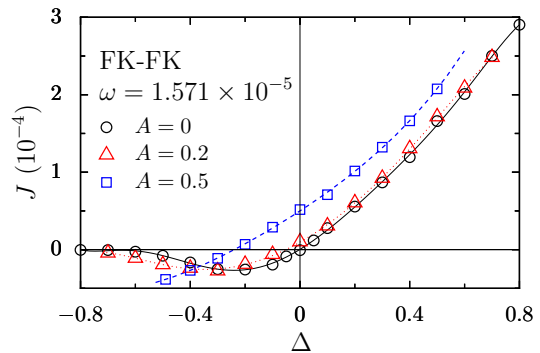}}
\caption{(color online). Action of a heat Brownian motor in two coupled
asymmetric FK-FK lattices. The heat baths are subjected to periodic
modulations in the form $T_{\text {L}}(t)=T_0[1+\Delta + A \; \hbox{sgn}(\sin{\omega
t})]$ and $T_{\text {R}}(t)=T_0(1-\Delta)$. The dimensionless reference
temperature is set as $T_0=0.09$ (see the Appendix \ref{app:units}
for the expressions of corresponding dimensionless units). Note that in distinct contrast to Fig. \ref{fig:jdelta} (d)
the ratchet effect now yields with a modulation strength $A\neq 0$ a nonvanishing  heat flow at zero bias $\Delta=0$.
Applying a substantial rocking strength, $A=0.5$, the
current bias characteristics can be manipulated as to inhibit  a negative differential thermal resistance (NDTR)
regime at negative bias values $\Delta$. Adapted from
\textcite{li2008EPL84}.}
\label{Fig-ShuttlingheatSecIII-ClassicalFKFK}
\end{figure}

A first possibility to introduce the necessary symmetry breaking is to use an
asymmetric material such as two coupled FK-FK lattices where two
segments possess different thermal properties
\cite{li2008EPL84,ren2010PRE81}, see in
Fig.~\ref{Fig-ShuttlingheatSecIII-ClassicalFKFK}. The directed heat
transport can be extracted  out of unbiased temperature fluctuations
by harvesting the static thermal rectification effect
\cite{li2004PRL93}.

Yet another possibility is to break the symmetry dynamically by
exploiting the nonlinear response induced by the harmonic mixing mechanism, stemming from a time-varying
two-mode modulation of the bath
temperature(s), i.e., $T_{\text{L(R)}}(t)=T_0[1\pm A_1
\cos(\omega t) \pm A_2 \cos(2\omega t+\varphi)]$. The second
harmonic driving term $A_2 \cos(2\omega t+\varphi)$ causes the intended
nonlinear frequency mixing \cite{li2009PRE80}.

In the adiabatic limit; i.e., if $\omega\rightarrow 0$, a nonzero
heat flux $J\neq 0$ can be obtained due to the nonlinearity of the heat conductance response.
The preferred direction of such heat flow is
determined by the intrinsic thermal diode properties. In contrast,
in the fast driving limit $\omega\rightarrow\infty$, the left- and
right-end of the system will essentially experience a time-averaged constant
temperature. This then mimics thermal equilibrium,
yielding $J\rightarrow 0$. Remarkably, by modulating the driving
frequency $\omega$ through the characteristic thermal response
frequency of the system, the intriguing phenomenon of a heat current {\it reversal}
can be observed \cite{li2008EPL84,li2009PRE80}. For this two segment system, an
optimal heat current can be obtained around this characteristic
frequency even when the two isothermal baths are oscillating
in synchrony with $T_{\text{L}}(t)=T_{\text{R}}(t)$, and the
current reversal can be realized by tuning the system size
\cite{ren2010PRE81}. In the harmonic mixing mechanism
the directed heat current is found to be proportional to the
third-order moment $\overline{(\Delta T(t)/2T_0)^3}$, i.e.,
$J\propto A^2_1 A_2 \cos{\varphi}$ \cite{li2009PRE80}. This enables
a more efficient way of manipulating  heat: the direction
of heat current can be reversed by merely adjusting the
relative phase shift $\varphi$ of the second harmonic driving.

Apart from using the FK-lattice as a source  of  nonlinearity, other
schemes of heat motors can be based on a Fermi-Pasta-Ulam (FPU) lattice structure,
a Lennard-Jones interaction potential \cite{li2009PRE80}, or also a Morse lattice structure
\cite{Gao2011ActaSinica60}.

Besides a manipulation  of bath temperatures, the shuttling of heat can  be
realized by use of a combination of  time-dependent mechanical control in otherwise symmetrical structures.
Depending on specific nonlinear system setups, an emerging  directed heat current can be
controlled by adjusting the relative phase among the acting drive forces  \cite{marathe2007PRE75}
or the driving frequency \cite{ai2010PRE81}. Multiple
resonance structures for the heat current  {\it vs.} the driving frequency of external forces can occur as well \cite{zhang2011PRE84}. It can be further demonstrated that for strict harmonic systems, periodic-force driving fails to shuttle heat. Moreover, it has been shown that even for anharmonic lattice segments composed of an additional energy depot, it is not possible to pump heat from a cold reservoir to a hot reservoir by merely applying external forces \cite{marathe2007PRE75,
zhang2011PRE84}.

It is  elucidative to compare these setups with the Feynman ratchet-and-pawl device of a heat pump \cite{hanggi2009RMP81,Broeck2006PRL96,
Broeck2008PRL100, Komatsu2006PRE73}. The latter is a consequence of
Onsager's reciprocal relation in the linear response regime:  if
a thermal bias generates a mechanical output, then an applied force
will direct a heat flow as a conjugate behavior. Therefore,
conjugated processes can be utilized for heat control as well. Other
well-known  thermally conjugated processes are the Seeback, Thomson and Peltier
effects in thermoelectrical devices, where the thermal bias induces
electrical currents, or vice versa. Recently, such a nanoscale magnetic
heat engine and pump has been investigated for a magneto-mechanical system,
which either operate as an engine via the application of a thermal bias to convert heat into useful work, or act as a cooler via applying magnetic fields or mechanical force fields to pump heat \cite{Bauer2010PRB81}.

\subsection{Quantum heat shuttling}

\begin{figure}
\centerline{\includegraphics[width=7cm]{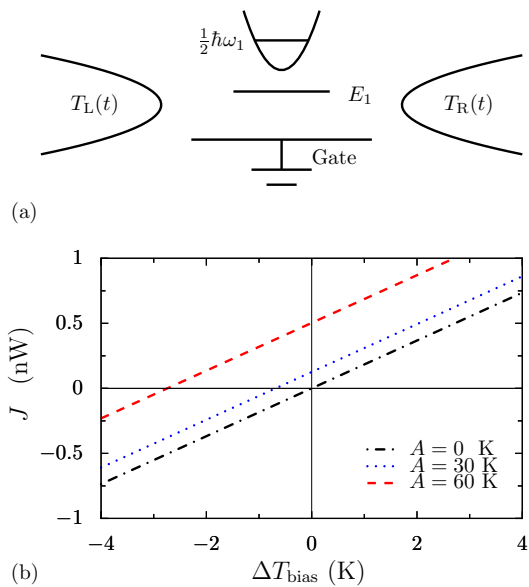}}
\caption{ (color online). Quantum heat shuttling. (a) Setup of
a  molecular junction acting as a heat shuttle. The short molecular
wire is composed of a single electronic level $E_1$ only that can be
gated and a single phonon mode at the fixed vibrational frequency
$\omega_1=1.4 \times 10^{14} s^{-1}$, typical for a carbon-carbon
bond.  The lead temperatures $T_{\text {L(R)}}(t)$ are subjected to
time-periodic modulations. (b) Quantum thermal rachet effect for
this molecular junction. Total directed heat current $J_{Q}$ as the
function of static thermal bias $\Delta T_{\text{bias}}$ for
different driving amplitude strengths $A$. The heat baths are
modulated as $T_{\text{L}}(t)=T_0 + \frac{1}{2}\Delta T_{\text{bias}} + A\cos{\Omega t}$ and $T_{\text {R}}(t)=T_0 - \frac{1}{2}\Delta T_{\text{bias}}$. The
reference temperature is set as $T_0=300$ K and the electronic wire
level is set as $E_1-\mu=0.138$ eV where $\mu$ is the chemical
potential. $J_Q$ is independent of the driving frequency $\Omega$ as a consequence of ballistic heat transfer in the adiabatic limit. After \textcite{zhan2009PRE80}.}
\label{Fig-ShuttlingheatSecIII-QuantumshuttlingSetup}
\end{figure}

The efficient shuttling of heat via a time-dependent modulation of
bath temperatures can be extended into the quantum regime when tunneling
and other quantum fluctuation effects come into play.
In clear contrast to the realm of electron shuttling \cite{hanggi2009RMP81,hanggi2002chemphys281,Galperin2007JPCM19,joachim2005PNAS102,remacle2005PNAS102}, however,
this aspect of shuttling quantum heat is presently still at an initial stage, although expected to undergo increasing future activity.

\subsubsection{Molecular wire setup}
In the following we consider one specific such case in some detail.
Let us consider a setup with a typical molecular wire for which the
heat transport is generally governed by both,
electrons and phonons. A schematic setup
based on a stylized molecular wire is sketched with
Fig.~\ref{Fig-ShuttlingheatSecIII-QuantumshuttlingSetup}(a)
\cite{zhan2009PRE80}. The single electronic level $E_1$ can be
conveniently modulated by a gate voltage and $\omega_1$ denotes the
vibrational frequency for a single phonon mode. The lead
temperatures $T_{\text{L(R)}}(t)$ undergo an adiabatically slow
periodic modulation for both electron and phonon reservoirs.  The latter
is experimentally accessible by use of a heating/cooling circulator
\cite{lee2005ACIE44}. In an adiabatic driving regime, the
asymptotic ballistic electron and phonon heat currents
$J^{\text{el(ph)}}_{Q}(t)$ can be calculated via  the celebrated
Landauer-like expressions \cite{Segal2003JCP119,Dubi2011RMP83}; i.e.,

\begin{eqnarray}
J^{\text{el}}_{Q}(t)=\int^{\infty}_{-\infty}\frac{d\varepsilon}{2\pi\hbar}(\varepsilon-\mu)
 \mathcal{T}^{\text{el}}(\varepsilon)
\left[f(\varepsilon,T_\text{L}(t))-f(\varepsilon,T_\text{R}(t))\right], \\
J^{\text{ph}}_{Q}(t)=\int^{\infty}_0
\frac{d\omega}{2\pi} \hbar\omega \mathcal{T}^{\text{ph}}(\omega)
\left[n(\omega,T_\text{L}(t))-n(\omega,T_\text{R}(t))\right]\;.\,\,\,\,
\end{eqnarray}
where $\mathcal{T}^{\text{el}}(\varepsilon)$ and
$\mathcal{T}^{\text{ph}}(\omega)$, respectively,  denote the temperature independent
transmission probability for electrons with energy $\varepsilon$
and phonons with angular frequency $\omega$. Here, the functions
$f(\epsilon,T_l(t))=[\text{exp}((\epsilon-\mu_l)/k_\text{B}T_l(t))+1]^{-1}$
and
$n(\omega,T_l(t))=[\text{exp}(\hbar\omega/k_\text{B}T_l(t))-1]^{-1}$
where $l=\text{L,R}$,  denote the Fermi-Dirac distribution and the
Bose-Einstein distribution, respectively. These functions both inherit a time-dependence which derives from the applied adiabatic
periodic temperature modulation.

Then, a finite total heat current $J_Q$ emerges, reading
\begin{equation}
J_Q=\overline{J^{\text{el}}_Q(t)+J^{\text{ph}}_Q(t)} \;.
\label{totalflow}
\end{equation}

This heat current $J_Q$ results as a consequence of the nonlinear dependence of quantum statistics on temperature, see in
Fig.~\ref{Fig-ShuttlingheatSecIII-QuantumshuttlingSetup}(b); -- note that in presence of modulation  $J_Q$
is nonzero even for vanishing thermal bias $\Delta T _{\rm{bias}}$.
This  ballistic heat transport is a pure quantum effect which will not occur in the classical diffusive
limit, being approached at ultrahigh temperatures. The efficient manipulation of heat
shuttling can be realized by applying the above-mentioned harmonic
mixing mechanism. Since the heat transport is carried also by
electrons, adjusting the gate voltage gives rise to a intriguing
control of heat current with the result that the direction of heat current
experiences multiple reversals.

The quantum heat shuttling of a dielectric molecular wire can also
be achieved upon periodically modulating the molecular levels while
this molecular wire is connected to two heat baths that are
characterized by distinct spectral properties \cite{segal2006PRE73}.
Interestingly, the pumping of quantum heat can be operated arbitrarily
close to the Carnot efficiency by a tailored stochastic modulation of the molecular levels
\cite{segal2008PRL101,segal2009JCP130}. Time-dependent phonon transport in the non-adiabatic regime and strong driving
perturbations has numerically been investigated by use of the nonequilibrium
Green's function (NEGF) approach \cite{Wang2008EPJB62,Dubi2011RMP83} for the case of coupled harmonic oscillator chains:
There, the coupling to the reservoirs is held at different temperatures with the latter switched on suddenly
\cite{cuansing2010PRB81}.

\subsubsection{Pumping heat via geometrical phase}

\begin{figure}
\centerline{\includegraphics[width=7cm]{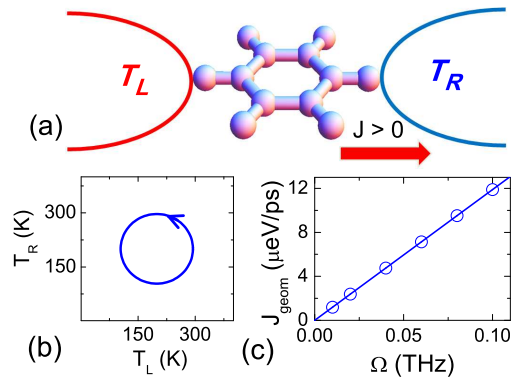}}
\caption{ (color online). Quantum heat pumping via a geometrical phase. (a) A schematic representation of a single molecular junction.
(b) Quantum heat transfer across the molecular junction is generated via
an adiabatic two-parameter variation of the left bath temperature $T_L (t)$ and the right bath temperature $T_R(t)$, which
maps onto a closed
(blue) circle. The arrow indicates the direction of the modulation protocol. (c) Geometrical phase
induced heat current, $J_{\mathrm{geom}}$, versus the angular modulation
frequency $\Omega$. The straight line corresponds to the analytic result
while the open circles denote the
simulation results. For further details we refer the reader to
\textcite{ren2010PRL104}.}
\label{Fig-ShuttlingheatSecIII-GeometricPhase}
\end{figure}

As discussed above,  a one-parameter modulation, e.g. via  the left-sided contact temperature  $T_L(t)$
is typically sufficient for quantum mechanical shuttling and rectification of heat, as sketched with Fig. \ref{Fig-ShuttlingheatSecIII-GeometricPhase}(a).


It should be noted, however, that a cyclic modulation involving at least {\it two} control parameters   generally induces  additional current contributions beyond its mere dynamic component. This is so because  with a variation
evolving in a (parameter) space of dimension ${\rm d}\ge2$ one typically generates geometrical properties (i.e. a nonvanishing, gauge  invariant curvature) in  the higher-dimensional parameter space of the governing dynamical laws for the observable,  in this case the heat flow. This geometrical properties  in turn  affect the resulting flow within an adiabatic  or even non-adiabatic parameter variation. With a cyclic adiabatic variation of multiple parameters this contribution to the emerging flow is thus of topological origin. It has been popularized in the literature under the label of a geometrical, finite curvature  or (Berry)-phase phenomenon \cite{Sinitsyn2009JPA42}. As a consequence, one  needs to consider the total heat flow  to be composed of two contributions, reading:

\begin{equation}
J_{\mathrm{tot}}= J_{\mathrm{dyn}}+J_{\mathrm{geom}}\;.
\end{equation}

Here, the geometrical contribution is proportional to the adiabatic small modulation frequency $\Omega$; implying that this correction is typically quite small in comparison with its non-vanishing  dynamical contribution. Therefore, to observe this component it is advantageous to use a two-parameter variation such that the dynamic component vanishes identically.

A sole geometrical contribution can be implemented, for example, by modulating the two bath temperatures in a way that the trajectory in the plane spanned by the two temperatures describes a circle, c.f.
Fig. \ref{Fig-ShuttlingheatSecIII-GeometricPhase}(b), -- in which case
there results a sole  Berry-phase-induced heat current , $J_{{\textup {geom}}}$, while its dynamical component, $J_{{\textup {dyn}}}$, is identically vanishing, $J_{{\textup {dyn}}}=0$ \cite{ren2010PRL104}. Also, different from the case of an irreversible, dynamical heat flux, this geometrical
contribution can be reversed upon simply reversing the protocol evolution.
The latter operation thus  provides a novel and convenient  method
for controlling energy transport. In \textcite{ren2010PRL104} the authors have demonstrated such nonvanishing quantum  heat pumping as the result of a nonvanishing geometrical phase. This situation is depicted  with  Fig. \ref{Fig-ShuttlingheatSecIII-GeometricPhase}(c).

A similar geometrical phase effect can also be present in classical setups;
e.g., for coupled harmonic oscillators in contact with Langevin
heat baths. In this case it has been demonstrated  that with a modulation of the two bath temperatures in time
the geometrical phase phenomenon emerges only for the higher order
moments of the heat flow; that is to say only beyond the average heat flux.
Only when nonlinearity or temperature-dependent parameters in an interacting system are present can
the geometrical phase  manifest itself in producing a nonvanishing heat current.

Moreover, the finite Berry-phase heat pump mechanism in both,
quantum and classical systems has been demonstrated to cause a breakdown of the so termed
``heat-flux fluctuation theorem'', the latter being valid for a
time-independent heat flux transfer. This fluctuation
theorem \cite{Saito2007PRL99, Campisi2011RMP83} can be restored only under special conditions in the
presence of a vanishing Berry curvature \textcite{ren2010PRL104}.

\subsection{Topological phonon Hall effect}

\begin{figure}\vspace{-3mm}
\centerline{\includegraphics[width=7cm]{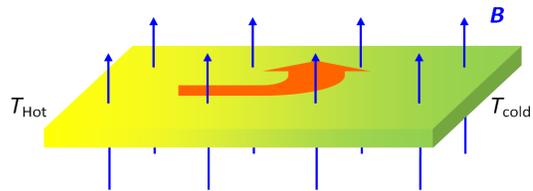}}\vspace{-7mm}
\caption{ (color online). A schematic illustration of the phonon Hall effect with heat flowing along the direction indicated by the (red) arrow.}
\label{Fig-ShuttlingheatSecIII-PHE}
\end{figure}

It is  known that a geometrical Berry phase yields profound effects on electronic transport properties in various Hall effect setups \cite{xiao2010RMP82}. Due to the very different nature of electrons and phonons, the phonon Hall effect (PHE) has been  discovered only recently in a paramagnetic dielectric \cite{strohm2005PRL95}, and subsequently confirmed by yet a different  experimental setup \cite{inyushkin2007JL86}. In particular one observes a transverse heat current  in the direction perpendicular to the applied magnetic field and to the longitudinal temperature gradient, see in Fig.~\ref{Fig-ShuttlingheatSecIII-PHE}. The discovery of this novel PHE renders the magnetic field to be another flexible degree of freedom for phonon manipulation towards the objective of energy and information control in phononics.

Since then, several theoretical explanations have been proposed
\cite{sheng2006PRL96, kagan2008PRL100, wang2009PRB80,
zhang2010NJP11} to understand the PHE by
considering the spin-phonon coupling. The spin-phonon coupling
has two possible origins: (i) It either derives from the magnetic
vector potential in ionic crystal lattices, where the vibration of
atoms with effective charges will experience the Lorentz force under
magnetic fields \cite{holz1972NCB9}, or (ii) it results from a Raman
(spin-orbit) interaction \cite{kornig1939PHYSICA6, vleck1940PR57,
orbach1961PRSLSA264, ioselevich1995PRB51}.
It has been shown that by introducing spin-phonon couplings, a ballistic system without nonlinear interaction even exhibits the possibility of thermal rectification \cite{zhang2010PRB81}.

Similar to the various Hall effects occurring for electrons, a topological explanation of the PHE has been provided in  \textcite{zhang2010PRL105, zhang2011JPCM23}. The heat flow in the PHE is ascribed to two separate contributions: the normal flow responsible for the longitudinal phonon transport, and the anomalous flow manifesting itself as the Hall effect of the transverse phonon transport. A general expression for the transverse phonon Hall conductivity is obtained in terms of the Berry curvature of phononic band structures. The associated topological Chern number (a quantized integer) for each phonon band is defined via integrating the Berry curvature over the first
Brillouin zone. For the two dimensional honeycomb and kagome-lattice, the authors observed phase transitions in the PHE, which correspond to the sudden change of the underlying band topology.
The  physical mechanism is rooted in
the touching and splitting of the phonon bands \cite{zhang2010PRL105, zhang2011JPCM23}.

Therefore, much alike for electrons in topological insulators  \cite{Hasan2010RMP82}, the  design of a family of novel phononic devices -- topological thermal insulators -- is  promising, with the bulk being an ordinary thermal insulator while the edge/surface constitutes an extraordinary thermal conductor.


\section{Summary, Sundries and Outlook} \label{Sec V}

With this Colloquium we took  the reader on a tour presenting the  state of the art of  the topic termed `Phononics', an emerging research direction which is expected to chime in the future with conventional `Electronics'. Particularly, we surveyed and  explained various  physical mechanisms that are exploited in devising the elementary phononic toolkit: namely,  a thermal diode/rectifier, thermal transistor, thermal logic gate and thermal memory. These building blocks for doing phononic electronics are rooted in the application of suitable static or dynamic control schemes for shuttling heat. We  further reviewed  recent attempts to realize such phononic devices which all are based on nanostructures and discussed first experimental realizations.

\subsection{Challenges}
In spite of the rapid developments of phononics in both science and technology, we should stress that this objective is still at its outset. To put these phononic devices to work, there remain lots of ambitious theoretical and severe experimental challenges to be overcome. For example, much  work is still required  for the physical realization of the phononic  toolbox and for entering the next stage of assembling operating networks that are both scalable and stable under ambient conditions.

\emph{Theory.}-- Interface thermal resistance. As we have pointed out the underlying mechanism for thermal rectifier/diode is based on an asymmetric interface thermal resistance. However, thus far a truly  comprehensive theory for this effect is lacking.

The current approaches for thermal transport across an interface, such as the acoustic mismatch (AMM) theory \cite{Little1959CanJPhys37} and the diffusive mismatch (DMM) theory \cite{Swartz1989RMP61}, are based on the assumption that phonon transport proceeds via a combination of either ballistic or diffusive transport on either side of the interface. Both schemes offer limited accuracy for nanoscale interfacial resistance predictions \cite{Stevens2005JHT127}, due to the neglect of the atomic details of actual interfaces. Specifically, the acoustic mismatch model assumes that phonons are transported across the interface without being scattered; i.e. they are ballistic while the diffuse mismatch model assumes the opposite, namely that the phonons are scattered diffusively. Thus, the effects of scattering on the interfacial thermal resistance act as upper and lower limits for  real situation.
In fact, both numerical and experimental studies in nanostructures ranging from nanotubes \cite{Zhang2005JCP123B, chang2008PRL101}, nanowires \cite{yang2010NT5}, to polyethylene nanofibers \cite{Henry2008PRL101, Henry2009PRB79, Shen2010NNano5} all show that phonons undergo anomalous diffusion - i.e. so termed  superdiffusion -,  being faster than normal diffusion but slower than ballistic transport.  Therefore, it is necessary to establish an improved theory describing thermal transport across the interface by taking into account the anomalous thermal transport characteristics of nanostructures.

Yet another  fact of  thermal rectification and negative differential thermal resistance is the role of (strong)  nonlinearity. How to set up a transport theory by incorporating nonlinearity in both, the quantum regime and the classical regime presents a challenge. The non-equilibrium Green's (NEGF) function method \cite{Wang2008EPJB62} certainly serves as an elegant mathematical framework. However, when  sophisticated phonon-phonon interactions (which derive from anharmonicity), and alike, become increasingly important (this being so  in realistic  nanostructures), the NEGF presents a cumbersome task. In the classical regime, an effective phonon theory has proven  useful in characterizing heat conduction \cite{Li2006EPL75, Li2007EPL78}.  Several classical studies have been advanced \cite{He2008PRE78, He2009PRB80, He2010PRE81}, which may as well be extended into the quantum regime.

\emph{Experiment.}--Experimental realization of practical phononic device depends on how to measure its thermal conductivity accurately. To this end, one needs to get rid of the contact thermal resistance. Thus far no well defined scheme capable to eradicate such contact thermal resistance has been put forward.

It also should be noted that presently the prime elementary phononic building block, i.e. the thermal rectifier, has been realized on  the micrometer scale \cite{Chang2006Sci314} and on the millimeter scale \cite{Kobayashi2009APL95, Sawaki2011APL98} only. The challenge for experimentalists is to make the sample smaller, for example towards a dozens of nanometers or even a few nanometers only. This then would boost the  rectification to much larger values and simultaneously would allow detecting the asymmetric interface thermal resistance.  The primary challenge ahead, however, is to validate the negative differential thermal resistance, i.e. the key element necessary in realizing the thermal transistor.

For electronic-like function a networking towards a  phononic-like logic operation is indispensable. In the event that this task could be achieved successfully  could then  facilitate the desirable operation of an all-phononic computer.


\subsection{Future prospects}

As emphasized above,  the area of phononics lingers still in its infancy, yet it is at the verge of blossoming up.\\

\emph{From phononics to acoustics, and vice versa.} --
As we elucidated  throughout this Colloquium the physical principle of phononic devices rests in the manipulation of phonon bands/spectra. This idea can be generalized to control any elastic/mechanical energy.  For example, inspired by the thermal diode, an acoustic diode has been proposed by \textcite{Liang2009PRL103} and subsequently realized experimentally by use of a nonlinear acoustic medium and phononic crystal \cite{Liang2010NATM9}, and a sonic crystal geometry \cite{Li2011PRL106}. \textcite{Boechler2011NM10} have demonstrated experimentally  elastic energy switching and rectification.  There is little doubt that  in  parallel to phononics, namely an acoustic transistor, logic gate  and possibly even a computer device may be realized in the foreseeable future. The ideas and concepts in controlling acoustic waves can as well be exploited for phononics. For example, phononic crystals have been demonstrated to be useful in manipulating acoustic wave propagation \cite{Liu2000Science289}. This concept has been extended recently to control heat flow on the nanoscale \cite{McGaughey2006PRB74,Landry2008PRB77,Yu2010NNano5, Hopkins2011NL11}. It would not come as a too big surprise when someday concepts such as a  ``heat-cloak'', ``super heat lens'', etc., come alive  via the implementation of phononic concepts by use of acoustic meta-materials \cite{Zhang2004APL85,Yang2004PRL93,Guenneau2007NJP9,Fang2006NMate5}. In fact,  active research is presently pursued  in controlling and manipulating mechanical/elastic energy which ranges from long wavelength elastic waves to very short wavelength thermal waves, a glimpse of this activity can be adapted from the Proceedings of the world first conference on `Phononics' \cite{Hussein2012AIPXX}.

\emph{PhoXonics: phononics plus photonics, and beyond.} -- Another promising prospect is based on the combination of phononics with photonics to control and manage  photon and phonon energy concomitantly  \cite{Maldovan2006APL88}. This synergy  might potentially enable one to use the solar energy more wisely.  For instance, a single-molecule phonon field-effect transistor has been designed wherein phonon conductance is controlled by a back-gate electric field \cite{Menezes2010PRB81}. Moreover, as temperature is the commonly applied control parameter for chemical and biological reactions, phononic devices may also find application for the local control of temperatures, for example in regulating molecular self-assembling processes with the  ``Lab-on-a-Chip'' technology.

\emph{Phononics and electronics.} -- Electrons carry as well heat. Therefore one can control the heat flow carried by electrons with the help of  electric/magnetic fields. For example, by asymmetrically coupling a quantum dot to its two leads one can construct a heat rectifier \cite{Scheibner2008NJP10}; -- upon applying a gate voltage it is possible to operate a heat transistor \cite{Saira2007PRL99}. These kind of devices combined with phononic circuits may then carry the potential  to manipulate dissipation of heat and cooling in nanoscale and molecular devices. Overall, there is also the potential that hybrid structures composed of electronic and phononic elements may lead to beneficial applications.

\emph{From heat conduction to radiation.}-- In this colloquium, our proposed phononic devices are based on the control of heat conduction, being assisted by lattice vibrations. A similar idea can be generalized to control heat radiation. Indeed,  Fan  and collaborators at Stanford  proposed a photon-mediated ``thermal rectifier through vacuum'' \cite{Otey2010PRL104}  which  makes constructive use of the temperature dependence of underlying electro-magnetic resonances. In the same spirit, Fan's group \cite{Zhu2011APLxxx} revealed  negative differential thermal conductance through vacuum; this in turn allows for the blueprint of a transistor for heat radiation.

Finally, let us end  with a celebrated quote by Winston Churchill:
``This is not the end. It is not even the beginning of the end.
But it is, perhaps, the end of the beginning.''


\begin{acknowledgments}
The authors would like to thank G. Casati for most insightful discussions and fruitful collaborations during the early stage of this endeavor. We are also  indebted  to  M. Peyrard for many useful discussions and valuable suggestions on this topic.  Moreover, we are grateful to Bambi Hu, Pawel Keblinski, Zonghua Liu, Tomaz Prozen, Peiqing Tong, Jiang-Sheng Wang, Jiao Wang, Chang-qin Wu, Hong Zhao and our group members at NUS,  Jie Chen, Jingwu Jiang, Jinghua Lan, Lihong Liang, Wei Chung Lo,  Xin Lu, Xiaoxi Ni, Lihong Shi, Bui Cong Tinh, Ziqian Wang, Gang Wu, Rongguo Xie, Xianfan Xu, Nuo Yang, Donglai Yao, Yong-Hong Yan,  Lifa Zhang, Sha Liu, Dan Liu, Xiang-ming Zhao, Kai-Wen Zhang, Guimei Zhu, Jiayi Wang and  John T. L. Thong, for fruitful collaborations in different stages of this project. We also like to thank J. D. Bodyfelt for carefully reading  parts of the manuscript and his helping comments.

This work has been supported by grants from Ministry of Education, Singapore, Science and Engineering Research Council, Singapore, National University of Singapore, and Asian Office of Aerospace R\&D (AOARD) of the US Air Force by grants, R-144-000-285-646, R-144-000-280-305, R-144-000-289-597, respectively; the start up fund from Tongji University (NL and BL), the National Natural Science
Foundation of China,  grant No.10874243 (L.W.), the Ministry of Science and Technology of China, Grant No. 2011CB933001 (G.Z.), and by the German Excellence
Initiative via the ``Nanosystems Initiative Munich'' (NIM) (P.H.) and also by the DFG priority
program DFG-1243 ``Quantum transport at the molecular scale'' (P.H.).
\end{acknowledgments}



\section*{Appendix: Nonlinear Lattice Models}
\appendix
\addtocounter{section}{1}
\addtocounter{equation}{-9}

\subsection{Lattice models}
\label{app:models}

In this Appendix we  introduce three archetype one-dimensional (1D) lattice models commonly used in the  investigation of heat transport. These are (i)  the linear Harmonic lattice, (ii) the nonlinear Fermi-Pasta-Ulam $\beta$ (FPU-$\beta$) lattice and (iii) the Frenkel-Kontorova (FK) lattice. The simplest harmonic lattice serves as the basic model from which the FPU-$\beta$ lattice and FK lattice are derived by complementing the dynamics with a nonlinear inter-atom interaction in the FPU-case and an on-site potential in the FK-case, respectively.

For a 1D harmonic lattice with $N$ atoms the  normal modes of the lattice vibrations are known as phonons.  The corresponding  harmonic lattice Hamiltonian explicitly reads
 \begin{equation}\label{Apdx1-Har-model-x}
 H=\sum_{i=1}^{N}\left[\frac{p^2_i}{2m}+\frac{k_0}{2}(x_i-x_{i-1}-a)^2\right],
\end{equation}
wherein the dynamical variables $p_i$ and $x_i$ $i=1,...,N$, denote the momentum and position
degrees of freedom for the $i$-th atom and  $x_0 \equiv x_1-a$.  The parameters $m$, $k_0$, $a$ denote the mass of the atom, the spring constant and the lattice constant, respectively.
The position variable $x_i$ can be replaced by the
displacement from equilibrium position as $\delta x_i= x_i -ia$ which we denote by the same symbol $ x_i \equiv  \delta x_i$ henceforth.
The Hamiltonian of Eq.~(\ref{Apdx1-Har-model-x}) with $ \delta x_0 = \delta x_1$ thus simplifies, reading:
\begin{equation}\label{Apdx1-Har-model}
H=\sum^{N}_{i=1}\left[\frac{p^2_i}{2m}+\frac{k_0}{2}\left(x_i-x_{i-1}\right)^2\right] \;.
\end{equation}
This form explicitly depicts the translational invariance of the free chain which implies momentum conservation.

Applying next  periodic boundary conditions $x_0\equiv x_{N}$,
the harmonic lattice of Eq.~(\ref{Apdx1-Har-model}) can be
decomposed into the sum of noninteracting normal modes (phonons)
with the dispersion relation reading
\begin{equation} 
\omega(q)=2\sqrt{\frac{k_0}{m}}\left|\sin(q/2)\right|,\,\,\,\,0\leq q \leq 2\pi \,
\end{equation}
where the continuous spectrum is due to the adoption of
thermodynamical limit $N\rightarrow\infty$. The harmonic lattice
possesses an acoustic phonon branch with  $\omega(q)\rightarrow 0$ as
$q\rightarrow 0$.

Although the 3D harmonic lattice model yields a satisfactory
explanation for the temperature dependence of experimentally
measured specific heat, it turns out that this very
model of noninteracting phonon modes fails to describe
a Fourier Law for heat transport. In
pioneering work  \cite{rieder1967JMP8} the authors proved
that for this 1D harmonic lattice the heat transport is
ballistic; this is due to the absence of phonon-phonon
interactions. In order to take the phonon-phonon interactions into
account, the harmonic chain model must be complemented with nonlinearity.

If a quartic inter-atom potential is added, one arrives at the
FPU-$\beta$ lattice with the corresponding  Hamiltonian reading:
\begin{equation}\label{Apdx1-FPU-model}
 H=\sum_{i=1}^{N}\left[\frac{p^2_i}{2m}+\frac{k_0}{2}(x_i-x_{i-1})^2+\frac{\beta_0}{4}(x_i-x_{i-1})^4\right],
\end{equation}
where the parameter $\beta_0$ is the nonlinear coupling strength.
Historically, the FPU-$\beta$ lattice has been put forward by Fermi, Pasta
and Ulam \cite{fermi1955} to study the issue of ergodicity  of a
nonlinear system dynamics. For some excellent reviews of the original FPU
problem we refer the readers to Refs.~\cite{ford1992PR213,berman2005Chaos15}). Surprisingly, the
FPU-$\beta$ lattice still fails to obey Fourier's law and the heat
conductivity $\kappa$ diverges with the system size proportional to
$\kappa\propto N^{\alpha}$ with $0<\alpha<1$
\cite{lepri1997PRL78}. This divergent behavior has recently been
experimentally verified for a system setup using quasi-1D nanotubes
\cite{chang2008PRL101}. The phonon modes of FPU-$\beta$ lattice are
also acoustic-like after re-normalization of the nonlinear part, due to
the conservation of total momentum.

If a periodic on-site substrate potential is added to the harmonic chain, one arrives at the FK lattice. Its Hamiltonian is given by
\begin{equation} 
 H=\sum_{i=1}^{N}\left[\frac{p^2_i}{2m}+\frac{k_0}{2}(x_i-x_{i-1})^2+\frac{V_0}{4\pi^2}\left(1-\cos{\frac{2\pi x_i}{a}}\right)\right],
\end{equation}
where parameter $V_0/4\pi^2$ denotes the nonlinear on-site coupling
strength. Here we  only consider the commensurate case where the
on-site potential assumes the same spatial periodicity as the harmonic
lattice. Notably this model with an on-site potential now breaks momentum conservation.
Among the various phenomenological models that mimic solid
state systems the FK model has been shown to provide a
suitable theoretical description for  possible nonlinear
phenomena such as the occurrence of commensurate-incommensurate phase
transitions \cite{floria1996AP45},  kink-like structures  and alike
\cite{braun2004,Braun1998Physrep306}. It has attracted
interest since it was first proposed by Frenkel and Kontorova
\cite{frenkel1938PhysZSowjetUnion13,frenkel1939JPhys-USSR1} in order to study
various surface phenomena. Recently it has been established that the FK lattice indeed does
exhibit normal heat conduction and thus obeys the Fourier law
\cite{hu1998PRE57}. This normal behavior is attributed to the
optical phonon mode where the phonon mode opens a gap as the momentum conservation is broken with the  on-site
potential.

\subsection{Local temperature and heat flow}
\label{app:units}
Dimensionless units constitute practical tools for the theoretical analysis and
numerical simulations. Here we provide a brief introduction to the dimensionless units used
in this report for the various lattice model setups.

Let us start with the simplest
lattice model of 1D Harmonic lattice of Eq.~(\ref{Apdx1-Har-model}).
For the harmonic lattice contacting a heat bath specified
by a temperature $T$, there are four independent parameters $m,a,k_0$
and $k_{\text{B}}$ where $k_{\text{B}}$ denotes the Boltzmann constant.
The dimensions of all the physical quantities that typically enter the issue of heat transport
can be expressed by the  proper combination of these four
independent parameters because there are only four fundamental
physical units involved: length, time, mass and temperature.

As a result, one can introduce the dimensionless variables by
measuring lengths in units of $[a]$, momenta in units of
$[a(mk_0)^{1/2}]$, temperature in units of $[k_0a^2/k_{\text{B}}]$,
frequencies in units of $[(k_0/m)^{1/2}]$, energies in units of
$[k_0a^2]$ and heat currents in units of $[a^2k^{3/2}_0/(2\pi
m^{1/2})]$. In particular, the Hamiltonian of Eq.~(\ref{Apdx1-Har-model})
can be transformed into a dimensionless form
if we implement the following substitutions:\\
\begin{equation}\label{Apdx2-subst}
H\rightarrow H  [k_0a^2], p_i\rightarrow p_i
[a(mk_0)^{1/2}], x_i\rightarrow x_i [a],
\end{equation}
where the so transformed dynamical variables   yield the
dimensionless variables to obtain
\begin{equation} 
H=\sum^{N}_{i=1}\left[\frac{p^2_i}{2}+\frac{1}{2}\left(x_i-x_{i-1}\right)^2\right],
\end{equation}

Typical physical values for atom chains are as follows: $a \sim 10^{-10}$ m,
$\omega_0\sim10^{13}$ sec$^{-1}$, $m \sim 10^{-26}- 10^{-27}$ kg,
$k_B=1.38\times10^{-23}$ J K$^{-1}$, we have
$[k_0a^2/k_{\text{B}}]\sim(10^2-10^3)$K.  This in turn implies that  room
temperature corresponds to a dimensionless temperature $T$ of the order  $0.1-1$ \cite{hu1998PRE57} .

To obtain the dimensionless FPU-$\beta$ lattice from Eq.~(\ref{Apdx1-FPU-model}), one cannot scale the five parameters
$k_{\text{B}}=a=m=k_0=\beta_0=1$ because one of them is redundant. Applying the
substitutions of Eq.~(\ref{Apdx2-subst}), we obtain the dimensionless form for the
FPU-$\beta$ Hamiltonian:
\begin{equation}\label{Apdx2-FPU-model-dml}
H=\sum^{N}_{i=1}\left[\frac{p^2_i}{2}+\frac{1}{2}\left(x_i-x_{i-1}\right)^2+\frac{\beta}{4}\left(x_i-x_{i-1}\right)^4\right]\,,
\end{equation}
with the dimensionless parameter  $\beta\equiv\beta_0 a^2/k_0$. It is
evident that the dimensionless nonlinear coupling strength $\beta$ is
generally not equal to unity. However, it can be shown that upon
adjusting $\beta$ becomes equivalent to vary the system energy or its
temperature.

The dimensionless FK Hamiltonian can also be obtained by use of Eq.~(\ref{Apdx2-subst}):
\begin{equation} 
H=\sum^{N}_{i=1}\left[\frac{p^2_i}{2}+\frac{1}{2}\left(x_i-x_{i-1}\right)^2+
       \frac{V}{4\pi^2}[1-\cos(2\pi x_i)]\right],
\end{equation}
where the dimensionless on-site coupling strength $V\equiv
V_0/k_0a^2$.

Thus far, we dealt with homogeneous lattice Hamiltonians. For thermal devices with more than one segment, each segment may possess its own  set of parameters such as a different spring constant or nonlinear coupling strength. In those cases, the reference parameter, for instance $k_0$, used to define a transformation in Eq.~(\ref{Apdx2-subst}) may be chosen to correspond to a natural parameter of the corresponding segment. In particular, the dimensionless Hamiltonian for each individual segment of a coupled FK-FK lattice may be written as
\begin{equation}\label{Apdx2-FKFK-model-dml}
H=\sum^{N}_{i=1}\left[\frac{p^2_i}{2}+\frac{k}{2}\left(x_i-x_{i-1}\right)^2+\frac{V}{4\pi^2}[1-\cos(2\pi
x_i)]\right],
\end{equation}
where $k$ is measured with the reference to a parameter $k_0$ which is  introduced {\it a priori}.

Next we discuss the results for expressing temperature and
heat current in dimensionless units. In our classical simulations  we typically used  Langevin thermostats with coupling the ``contact'' or end particles to the heat baths at their corresponding temperature. More precisely, one adds to the corresponding Newtonian equation of motion a Langevin fluctuating term  which satisfies the fluctuation-dissipation relation, e.g. see in \textcite{Hanggi1990RMP62}. Towards this goal we made use of the equipartition theorem of classical statistical mechanics to define, for example,  the local temperature $T_i$ via its average atomic kinetic energy; i.e.,
\begin{equation}
k_{\text{B}}T_i\equiv \left<\frac{p^2_i}{m}\right>\rightarrow
T_i\equiv \left<p^2_i\right>,
\end{equation}
where the arrow indicates the dimensionless substitution
$T_i\rightarrow T_i [k_0a^2/k_{\text{B}}]$ and $p_i\rightarrow
p_i [a(mk_0)^{1/2}]$, and $\left<\cdot\cdot\cdot\right>$ denotes the (long)
time-average or, equivalently, its ensemble-average in numerical simulations, thus implicitly assuming ergodicity (in mean value).

Unlike temperature, the expression for the heat current is model
dependent. To arrive at a compact  expression  we first rewrite the 1D
lattice Hamiltonian in the more general form:
\begin{equation}
H=\sum_{i}\left[\frac{p^2_i}{2}+V(x_{i-1},x_i)+U(x_i)\right],
\end{equation}
where $V(x_{i-1},x_i)$ denotes the inter-atom potential and $U(x_i)$
is the on-site potential. As a result of the continuity equation for
local energy, the local, momentary  heat current can be expressed as \cite{Lepri2003PR377}:
\begin{equation}
J_i= - \dot{x}_i\frac{\partial{V(x_{i-1},x_{i})}}{\partial{x_i}}.
\end{equation}
Consequently, the expression of heat current depends only on the
form of the inter-atom potential $V(x_{i-1},x_{i})$. One should
notice that although the on-site potential $U(x_i)$ does not enter
into the expression of heat current explicitly, it does, however, influence the
heat current implicitly through the dynamical equations of motions.

The heat currents itself are again obtained via the time average over an extended time span.  For steady state setups with fixed
bath temperatures the resulting heat currents are time-independent and, as well, independent of the particular site index ($i$) within the particular chain segment. Likewise, with periodically varying bath temperatures $T(t)$, the resulting ensemble average is also
time-periodic; an additional time-average over the temporal period of  $T(t)$ yields the cycle-averaged, time-independent
heat flux. Alternatively, an explicit long time average again produces this very time-independent value for the heat current.

\subsection{Power spectra of FPU-$\beta$ and FK lattices}
\label{app:power}

The power spectrum (or power spectral density) describes the distribution of a system's
kinetic energy falling within given frequency intervals. For a homogeneous  lattice composed of identical particles
the velocity  $v_{i}(t) \equiv v(t)$ of a particle located at site $i$  becomes independent of position $i$; the power spectrum then can be conveniently calculated by the Fourier
transform of the corresponding velocity degree of freedom to yield:
\begin{equation} 
P(\omega) = \left|\lim_{t_0 \rightarrow \infty} \frac 1{t_0}
\int_0^{t_0} v(t) e^{-i \omega t} dt\right|^2.
\end{equation}

In doing so, the power spectrum of the FPU-$\beta$ model of Eq.~(\ref{Apdx2-FPU-model-dml})
depends on the temperature. In the low temperature regime
the FPU-$\beta$ dynamics is close to a harmonic lattice, yielding $0<\omega<2$.
In contrast, in the high temperature regime it is the anharmonic part that starts to dominate.
In this latter regime an
approximate theoretical estimate then yields $0<\omega<C_0 (T\beta)^{1/4}$,
with $C_0=2\sqrt{2\pi}\Gamma (3/4)3^{1/4}/\Gamma(1/4)\approx
2.23$, where $\Gamma$ denotes the Gamma function \cite{li2005PRL95}.
Therefore, upon increasing the  temperature then causes a rightward shift of the  power spectrum
towards higher frequencies.
The Parseval's theorem then dictates
that the area below the curve is proportional to the average
kinetic energy of the particle; i.e.,
$\int_0^{\infty}P(\omega)d\omega \sim \langle E_{kin}\rangle$.\\

\begin{figure}
\includegraphics[width=\columnwidth]{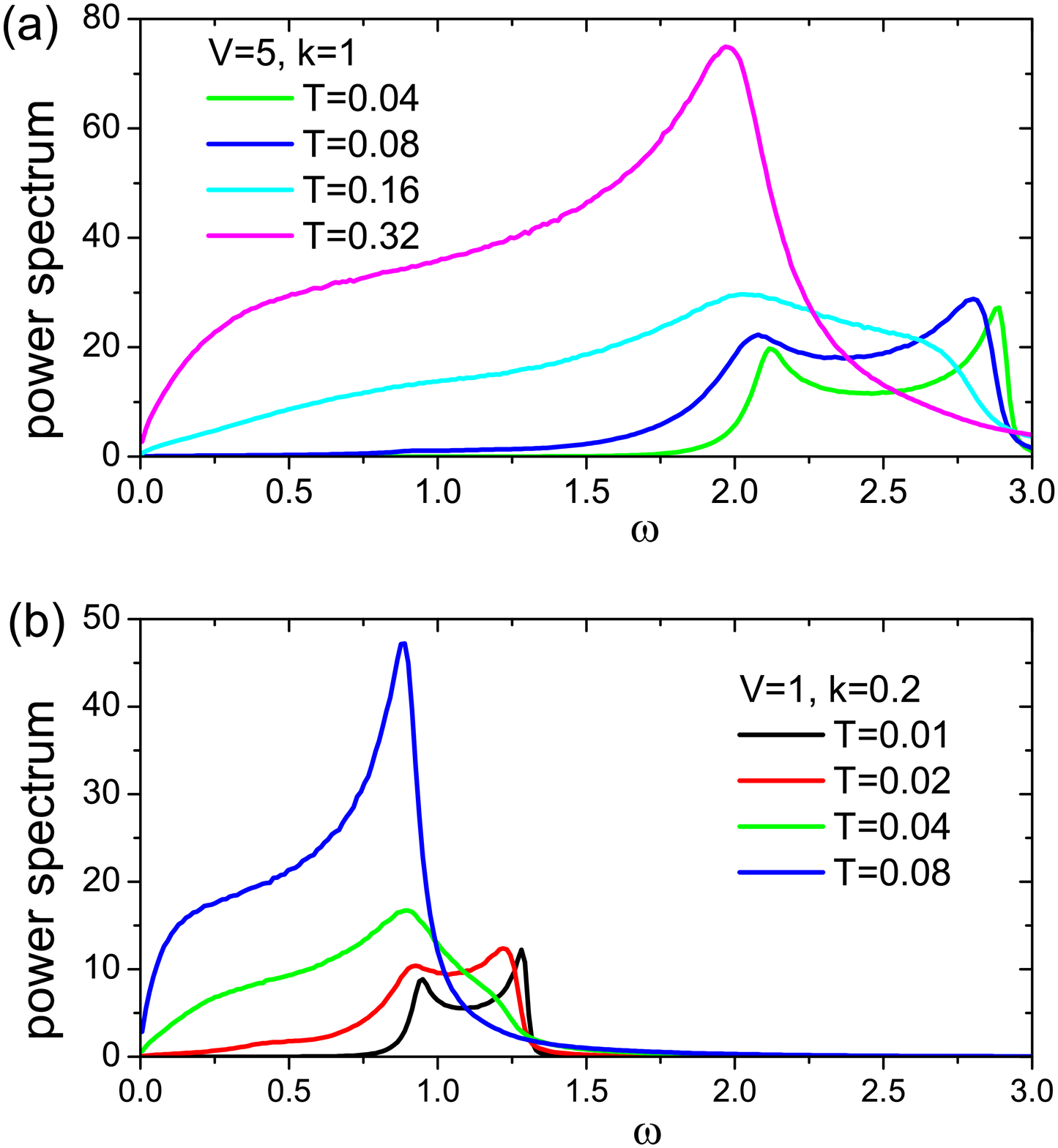}
\vspace{-0.5cm} \caption{\label{fig:FKspectrum} (color online). Temperature dependent power spectra. The variation of  the power spectrum at different temperatures {\it vs.} the angular frequency $\omega$ (both in dimensionless units) in an FK-lattice with 10000 sites for two different sets of parameters in (a) and (b). The features of these nonlinear  FK-power spectra  provide the seed for  the {\it modus operandi} in a thermal diode setup as discussed in Sec \ref{II-A-1}.  The  panel  (a) is for a FK-coupling strength of V=5 and a strength for the  spring constant of k=1; panel (b) is for a coupling strength V=1 and a spring constant value  set at k=0.2.}
\end{figure}

For the FK model in  Eq.~(\ref{Apdx2-FKFK-model-dml}) the form of the power spectrum depends sensitively  on temperature.
In the low temperature limit, the atoms are confined in the
valley of the on-site potential. Upon linearizing Eq.~(\ref{Apdx2-FKFK-model-dml}) the phonon band can be  extracted to read:
\begin{equation}
\sqrt{V}<\omega <\sqrt{V+4k}.  
\end{equation}

In contrast, in the high temperature limit the on-site potential can be
neglected; thus the system dynamics becomes effectively reduced to a harmonic chain dynamics, whose phonon band extends to:
\begin{equation}
0<\omega<2\sqrt{k}.  
\end{equation}
The crossover  temperature $T_c$ can be approximated as:
$T_{cr}\approx V/(2\pi)^2$. Its value depends on the height of
on-site potential. This in turn implies different values for the two   segments of the
thermal diode setup, see Fig.~\ref{fig:FKspectrum}. This difference is at the heart  of the thermal rectifying mechanism.


\bibliographystyle{apsrmplong}

\end{document}